\documentclass{aa}  
\usepackage{graphicx}
\usepackage{txfonts}
\usepackage{amsmath}
\usepackage[colorlinks,allcolors=blue,bookmarks=false,hypertexnames=true]{hyperref} 
\usepackage{xcolor}
\usepackage{placeins}

\begin{document} 

\title{Scale-dependent surface and volume density properties of filaments in molecular clouds}

\author{Guo-Yin Zhang\inst{1}
       \and
       Alexander Men'shchikov\inst{2}
       \and
       Jin-Zeng Li\inst{1}
       }
          
\institute{National Astronomical Observatories, Chinese Academy of Sciences, A20 Datun Road, Chaoyang District, Beijing 100101, China\\
      \email{zgyin@nao.cas.cn}
      \and 
      Universit\'{e} Paris-Saclay, Universit\'{e} Paris Cit\'{e}, CEA, CNRS, AIM, 91191, Gif-sur-Yvette, France\\
      \email{alexander.menshchikov@cea.fr}
     }
\date{Received March 23, 2026; accepted April 30, 2026}
\titlerunning{Scale-dependent properties of filaments in molecular clouds}
\authorrunning{G.-Y. Zhang, A. Men'shchikov, and J.-Z. Li}

\abstract
{
We present a systematic analysis of scale-dependent properties of filamentary structures in seven nearby molecular clouds -- \object{Taurus}, \object{Ophiuchus}, \object{Perseus}, \object{Orion A}, \object{California}, \object{IC 5146}, and \object{Vela C} -- using the multiscale extraction method \textit{getsf}. Alongside the usual surface density profiles $\Sigma(r)$, we derived volume density profiles $\rho(r)$ for a large sample of filaments, providing new observational constraints on their three-dimensional structure. The half-maximum widths $H$ and $h$ of the surface and volume density profiles, respectively, systematically increase with the spatial scale, following power laws $\tilde{H} \propto Y^{0.50}$ and $\tilde{h} \propto Y^{0.37}$, with distributions spanning $\sim$\,0.01--1\,pc across all scales, challenging the notion of a universal filament width of $\sim$\,0.1\,pc. The median volume density slopes $\tilde{\beta} \approx 2.1$--$2.4$ are systematically lower than the value $\beta = 4$ expected for an isothermal cylinder in hydrostatic equilibrium. For shallow profiles with $\beta \lesssim 1$, the volume density width $h$ falls below the surface density width $H$ by one to two orders of magnitude, demonstrating that surface density widths overestimate the true physical extent of filaments with shallow profiles. Volume density contrasts are substantially higher than surface density contrasts ($\tilde{C}_{\rho} \approx 17$--$52$ versus $\tilde{C}_{\Sigma} \approx 1.1$--$2.7$), confirming that filaments are substantially more prominent in three dimensions than their projected appearance suggests. The median linear densities of filaments increase linearly with the spatial scale, $\tilde{\Lambda} \propto Y$, with the fraction of supercritical filaments ($\Lambda > 15\,M_\odot$\,pc$^{-1}$) increasing strongly with the scale and varying widely among clouds, from $\sim$\,7\% in Taurus and Ophiuchus to $\sim$\,54\% in Vela C, which is broadly consistent with the known star formation activity of the clouds. The statistical properties of the combined filament linear density function across all seven clouds will be presented in a subsequent paper. Measured filament widths and slopes systematically depend on the angular resolution and distance, highlighting the importance of accounting for resolution bias in comparative filament studies.
}  


\maketitle

\section{Introduction}
\label{introduction}

Filamentary structures in the cold interstellar medium (ISM) play a central role in star formation \citep[] [for recent reviews]{Hacar+2023, Pineda+2023}. \textit{Herschel}\footnote{\textit{Herschel} is an ESA space observatory with science instruments provided by the European-led Principal Investigator consortia and with important participation from NASA \citep{Pilbratt+2010}.} imaging surveys of nearby Galactic clouds have revealed that filaments dominate the mass budget of molecular clouds at high densities and host most prestellar cores and star formation \citep{Andre+2010, Molinari+2010, Schisano+2014, Konyves+2015, Marsh+2016, DiFrancesco+2020}.

A notable result from \textit{Herschel} studies is the apparently common width (${\sim\,}0.1$ pc) of filaments observed in various molecular clouds \citep{Arzoumanian+2011, Arzoumanian+2019}. Similar widths were found in molecular clouds such as Taurus, Ophiuchus, and IC\,5146, appearing largely independent of filament surface density or environmental conditions \citep{Palmeirim+2013, Andre+2014, Arzoumanian+2019}. This characteristic width may represent a fundamental property associated with the physics underlying filament formation and evolution, potentially linked to the magnetized turbulent correlation length within molecular clouds \citep{Andre+2022}. Recent work for Ophiuchus indicates, however, that while the median width is consistent with this typical value, it represents a cloud-averaged property with a broad distribution \citep{Jia+2025}. Early observational evidence for a true cylindrical geometry was provided by \cite{Li+Goldsmith2012}, who used multiple HC$_3$N transitions toward the B213 filament in Taurus to derive a line-of-sight dimension comparable to its projected width, supporting the view that filaments are finite structures with a well-defined outer boundary, consistent with the finite-cylinder model developed in \cite{Menshchikov+Zhang2026subm}.

However, the interpretation of these observed widths faces both observational and methodological challenges. Recent studies have raised important questions regarding this universal width, suggesting that measured filament widths could be significantly affected by the insufficient angular resolution of \textit{Herschel} observations \citep{Panopoulou+2017, Panopoulou+2022}, similar to distance effects reported for observed sources \citep{Louvet+2021}. The observed correlation between filament widths and distances to molecular clouds \citep{Panopoulou+2022} challenges the concept of a universal intrinsic scale for filaments. Although convergence tests by \cite{Andre+2022} confirmed nearly constant widths for filaments in nearby star-forming regions, suggesting that resolution effects may not fully explain the observed width distribution, the issue remains contentious.

Beyond resolution effects, filament properties derived in previous studies may be affected by systematic methodological biases. \cite{Menshchikov+Zhang2026subm} demonstrate that the traditional approach of fitting surface density profiles with the Plummer function is fundamentally inconsistent with the finite nature of filamentary structures, and they show that deriving filament widths and slopes by fitting median surface density profiles can lead to large systematic errors. Furthermore, they find that the standard assumptions relating volume and surface density slopes via $\beta=\gamma+1$ and assuming similar widths ($h\approx H$) are generally invalid for shallow profiles ($\beta\lesssim 2$). To address these issues, we developed a fitting method that accounts for the finite nature of filaments, providing an analytical fitting function for surface density profiles and enabling the consistent derivation of physical volume density parameters.

The formation of filamentary structures involves a complex interplay between turbulence, magnetic fields, and gravity \citep[e.g.,] []{Fischera+Martin2012, Hennebelle2013, Hennebelle+Andre2013, Federrath2016, Federrath+2021}. Magnetohydrodynamics (MHD) simulations consistently produce filaments as natural outcomes of turbulent ISM evolution, with widths influenced by gravitational collapse and magnetic support \citep{Seifried+2017, Ntormousi+Hennebelle2019}.

Observational studies have revealed hierarchical substructuring of filamentary structures, with larger-scale filaments containing intertwined smaller-scale sub-filaments or fibers \citep[e.g.,] []{Hacar+2013, Hacar+2018, FernandezLopez+2014, Shimajiri+2019}. These findings challenge simplistic models of uniform cylindrical filaments and have implications for both filament formation theories and fragmentation processes leading to star formation, potentially influencing the resulting core mass function (CMF) and 
stellar initial mass function \citep{Motte+1998, Alves+2007, Konyves+2015, Konyves+2020, Louvet+2021}. Despite this progress, fundamental questions remain concerning the physical mechanisms controlling filament widths, the role of filament substructures in star formation, and the impact of finite angular resolution on derived filament properties.

Addressing these questions requires the systematic study of filaments and sub-filaments (fibers) detected at distinct spatial scales. The multiscale, multiwavelength extraction method \textit{getsf} \citep{Menshchikov2021method} is specifically designed to separate sources from filaments, subtract filament backgrounds, and detect filaments at arbitrary spatial scales while automatically producing their measurements. Critically, \textit{getsf} has no free parameters, eliminating subjective decisions in parameter selection that affect other methods. The performance of \textit{getsf} has been carefully evaluated in detailed benchmarks \citep{Men'shchikov2021bench}, demonstrating its reliability for systematic studies of filaments and cores.

In this work, we explore filamentary structures in \textit{Herschel} images of well-known star-forming regions at distances of 140 to 920 pc across a wide range of spatial scales from 14 to 216{\arcsec} using the advanced multiscale approach enabled by \textit{getsf}. By explicitly examining the scale dependence of filament widths, radial profiles, and slopes using the improved fitting method of \cite{Menshchikov+Zhang2026subm}, we aim to clarify the underlying physical processes controlling filament formation and evolution. We also assess potential biases introduced by observational resolution, thereby enhancing the reliability of derived filament properties and providing new constraints on the universality of filament widths. The statistical properties of the combined filament linear density function across all seven clouds will be presented in a subsequent paper. We describe the observational data and derived surface density images in Sect.~\ref{obs}, present our multiscale method of filament extraction in Sect.~\ref{approach}, describe the results of our analysis in Sect.~\ref{results}, discuss the results in Sect.~\ref{discuss}, and summarize our work in Sect.~\ref{conclusions}.

\begin{figure*}[ht!]
\includegraphics[width=1.0\hsize]{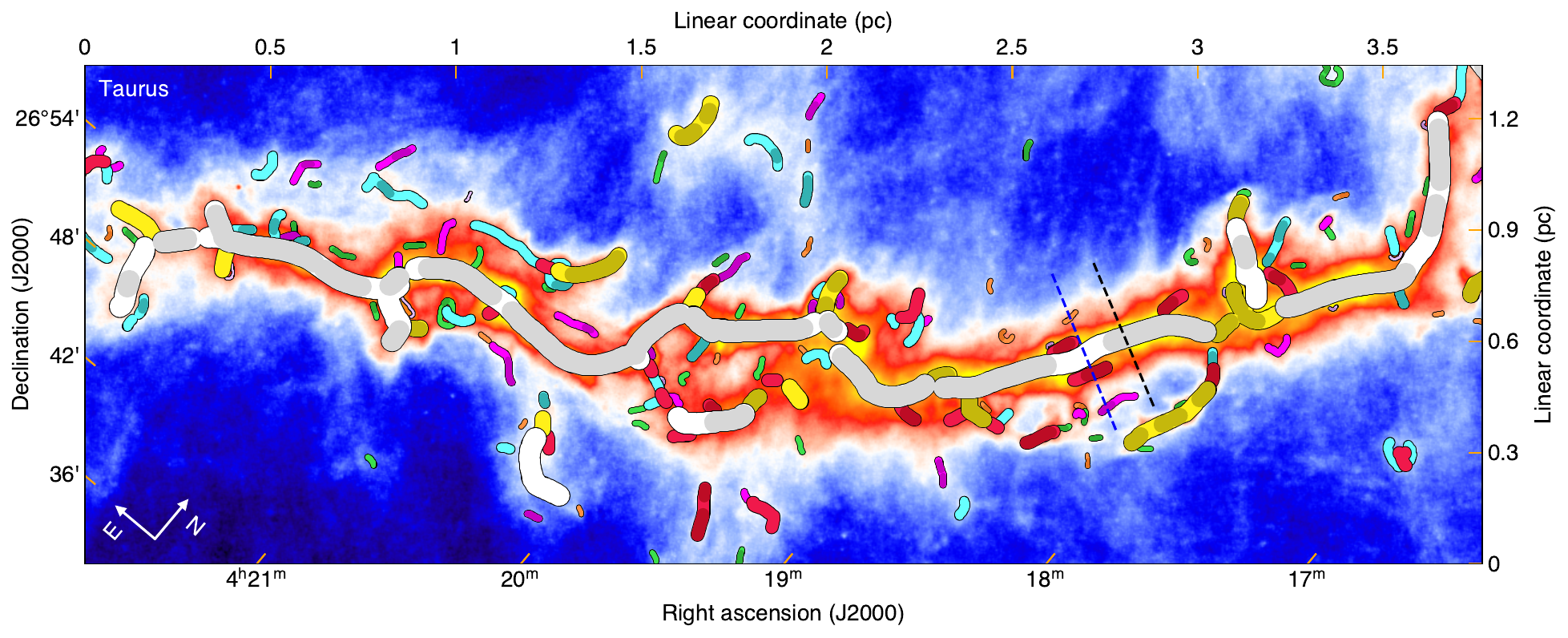}
\includegraphics[width=1.0\hsize]{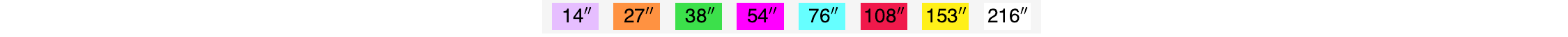}
\caption{
Zoomed-in surface density image of the \object{Taurus} molecular cloud, overlaid with scale-dependent skeletons traced by \textit{getsf} on scales of 14--216{\arcsec}. Widths of the colored skeletons are proportional to the scale (specified in the color bar) on which the filaments are detected. Darker skeleton colors indicate filament segments with acceptably good radial profiles (Sect.~\ref{fil_select}). Two dashed lines indicate the filament cuts shown in Fig.~\ref{fig:taurus_cut_profiles}.
}
\label{fig:filament.skeleton.taurus}
\end{figure*}

\section{Observed images and derived surface densities}
\label{obs}

\subsection{Observational data}

We used archival \textit{Herschel} data\footnote{\url{http://archives.esac.esa.int/hsa/whsa/}} for seven nearby molecular clouds: \object{Taurus}, \object{Ophiuchus}, \object{Perseus}, \object{Orion A}, \object{California}, \object{IC 5146}, and \object{Vela C}. The data comprise images in five \textit{Herschel} wavebands at 70, 160, 250, 350, and 500 $\mu$m from the HOBYS and Gould Belt surveys \citep{Andre+2010, Motte+2010, Harvey+2013}. The half-power beam width (HPBW) resolutions of these images are 8.4, 13.5, 18.2, 24.9, and 36.3{\arcsec}, respectively. We resampled all images to a common 3{\arcsec} pixel size using \textit{swarp} \citep{Bertin+2002}. We adopted distances of 140, 144, 294, 432, 470, 760, and 920 pc for \object{Taurus}, \object{Ophiuchus}, \object{Perseus}, \object{Orion\,A}, \object{California}, \object{IC\,5146}, and \object{Vela\,C}, respectively. All distances were derived using a uniform Bayesian method combining \textit{Gaia} DR2 parallaxes with stellar photometry, enabling reliable cross-cloud comparisons \citep{Zucker+2019,Zucker+2020}.

These seven clouds were selected because they span a wide range of distances (140--920\,pc), star formation activities, and environmental conditions, while having high-quality archival \textit{Herschel} multiband data available from the HOBYS and Gould Belt surveys. \object{Taurus} and \object{Ophiuchus} are nearby, low-mass star-forming regions with relatively quiescent activity \citep{Kenyon+2008,Wilking+Allen2008}. \object{Perseus} and \object{California} represent intermediate cases: \object{Perseus} is actively forming low- to intermediate-mass stars \citep{Bally+2008}, while \object{California} is a giant molecular cloud (GMC) with a surprisingly low star formation rate \citep{Lada+2009}. \object{Orion\,A} is the nearest region of high-mass star formation, hosting
the formation of both low- and high-mass stars \citep{Meingast+2016}, and \object{Vela\,C} is a GMC with active low- and intermediate- and high-mass star formation \citep{Massi+2019}. \object{IC\,5146} provides an example of a moderately active cloud with a prominent filamentary network \citep{Johnstone+2017}. Together, these clouds offer a representative cross-section of conditions in nearby star-forming regions.

\subsection{Surface density maps}

We derived H$_2$ surface density maps for each molecular cloud using the \textit{hires} algorithm \citep{Menshchikov2021method}. The method fits the spectral energy distribution of each pixel with a modified blackbody, assuming optically thin emission, a dust opacity law $\kappa_\nu \propto \nu^2$ \citep{Hildebrand1983,Ossenkopf+Henning1994}, and a
dust-to-gas mass ratio of 0.01. This opacity law is consistent with that adopted in previous studies of the same clouds, enabling direct comparisons. The fitting was performed using the \textit{fitfluxes} utility \citep{Menshchikov2016}.

To ensure accurate absolute calibration, we computed large-scale zero-level offsets for the \textit{Herschel} images by comparison with \textit{Planck} images \citep{PlanckCollaboration+2014} at corresponding wavelengths \citep{Bernard+2010}. By combining spatial information from the \textit{Herschel} images at different angular resolutions, we obtained high-resolution maps of H$_{2}$ surface density (and dust temperature) in each cloud (Figs.~\ref{fig:filament.skeleton.clouds.a}--\ref{fig:filament.skeleton.clouds.c}) with an effective angular resolution of 13.5{\arcsec}. A detailed description of the \textit{hires} method is given in \cite{Menshchikov2021method}.

\subsection{Validation}

We performed two consistency checks to validate the derived surface densities. First, we verified the accuracy of the zero-level offsets by comparing dust temperatures derived from fitting adjacent wavelength pairs: \{160, 250\}, \{250, 350\}, and \{350, 500\}\,$\mu$m. The median temperature differences between these pairs were below 1\%, confirming the consistency of the zero-level offset calibration.

Second, we verified the consistency between high- and low-resolution surface densities by degrading the 13.5{\arcsec} resolution images to 36.3{\arcsec} and computing their relative differences. The median relative differences were 4.2, 1.4, 2.6, 2.0, 2.9, 3.2, and 0.94\% for \object{Taurus}, \object{Ophiuchus}, \object{Perseus}, \object{Orion\,A}, \object{California}, \object{IC\,5146}, and \object{Vela\,C}, respectively. These small differences confirm that the \textit{hires} method accurately preserves surface densities while enhancing the effective angular resolution. The absolute accuracy of the method has been validated against simulated data in \cite{Menshchikov2021method}.

\section{Multiscale filament extraction}
\label{approach}

In contrast to sources, filament detection is fundamentally scale-dependent, and a single skeleton appropriate for one spatial scale cannot fully describe the complexity of observed multiscale, highly substructured filaments. Resolved filaments often appear composed of thinner sub-filaments (fibers) on smaller scales down to the angular resolution, with widths, profiles, and crest intensities varying considerably along their skeletons.

The \textit{getsf} method\footnote{\url{http://irfu.cea.fr/Pisp/alexander.menshchikov/}} is capable of detecting filaments at various spatial scales. Previously applied methods traced filament crests directly in observed or derived images, such as surface density maps \citep[e.g.,] []{Arzoumanian+2011,Palmeirim+2013,Arzoumanian+2019}. This approach inevitably reduces the accuracy of skeleton derivation because observed images contain fluctuations on all spatial scales, from the smallest (of order the angular resolution) to the largest (of order the cloud or image size, whichever is smaller). Small-scale fluctuations create spurious gradients and crests completely unrelated to the filamentary structures of interest.

Filaments in observed star-forming regions have characteristic physical widths (transverse scales), and their detection in \textit{Herschel} images containing structures and strong fluctuations on all spatial scales is severely hampered without prior scale separation. Indeed, the \textit{disperse} method \citep{Sousbie2011}, routinely used in filament studies, produces inaccurate skeletons with spurious wiggles for wide (resolved) filaments (Fig.~\ref{fig:disperse_overlay}) because \textit{disperse} is strongly affected by small-scale background and noise fluctuations. Similar problems affect other scale-independent filament detection methods \citep[e.g.,] []{Schisano+2014}.

\subsection{Maximum sizes of sources and filaments}
\label{maxsizes}

The maximum sizes $X$ and $Y$ of sources and filaments, respectively, are the single input parameter of \textit{getsf} that must be estimated before extraction \citep[Sect.~3.1.3 in] []{Menshchikov2021method}. This parameter defines the maximum scale $S_{\!J}$ for image decomposition, excluding unnecessarily large scales from processing. For example, the images studied in this work have largest scales of $S_{\!J} \sim 10^{4}${\arcsec} (of order the image size), but visual inspection shows that the structures of interest are limited to scales $S_{\!j} \lesssim 200${\arcsec} ($j=1, 2, \dots, J$, with $J = 99$). With $Y_{K} = 216${\arcsec}, the spatial decomposition therefore adopts an upper limit $S_{\!J} = 4 \max (X,Y) \approx 864${\arcsec}, which is sufficient to encompass all structures of interest.

The maximum size of filaments ($Y$) is more important than that of sources ($X$) because the derived filament backgrounds depend directly on the adopted value of $Y$. Source backgrounds do not depend on $X$; they are obtained by multidirectional interpolation of pixel values just outside the source footprints. However, filament backgrounds are far more uncertain and cannot be determined accurately by interpolation alone. The approach taken by \textit{getsf} \citep[Sect.~3.2 in] []{Menshchikov2021method} is to derive filament backgrounds by removing spatial scales with $S_{\!j} < 4 Y$.

We used a uniform maximum source size $X = 90${\arcsec} in each \textit{getsf} run. To extract scale-dependent filaments and estimate their backgrounds, we defined a set of maximum filament sizes $Y_{k} =$ 14, 27, 38, 54, 76, 108, 153, and 216{\arcsec} ($k=1, 2, \dots, K$, with $K = 8$), with values from 27{\arcsec} onward spaced by a factor of $2^{1/2}$. We extracted sources and filaments in each molecular cloud using $Y_{K} = 216${\arcsec} and executed additional shorter runs to produce backgrounds, skeletons, and measurements for the remaining $Y_{k}$ values from 153 down to 14{\arcsec} (see Figs.~\ref{fig:filament.skeleton.taurus}, \ref{fig:taurus_cut_profiles}).

\subsection{Tracing filaments with unique skeletons}
\label{trace_unique}

Scale-dependent skeletons of the \textit{getsf} method trace filaments effectively because signals on scales much smaller or larger than $S_{\!j}$ are removed from the images \citep[Sect.~3.4.5 in] []{Menshchikov2021method}. Therefore, filaments with matching half-maximum widths $H \approx S_{\!j}$ become the most prominent structures on spatial scale $S_{\!j}$, where $S_{\!j}$ is the full width at half maximum of the Gaussian kernel used in the image decomposition \citep{Menshchikov2021method}. However, sufficiently dense filaments are often detectable within a fairly large range of scales, and hence the same filament can be traced with skeletons on several adjacent scales. For example, a simple model of an isolated straight filament would be detectable on all spatial scales, but all single-scale skeletons would trace the crest of the same filament. To ensure that each filament with a certain width $H$ is traced by a unique skeleton on an appropriate scale $S_{\!j}$, the scale-dependent skeletons must be further analyzed and cleaned.

To eliminate redundant skeleton points, \textit{getsf} iterates over each spatial scale $k$ from the angular resolution $O$ to the maximum filament size $Y_{K}$. On each scale $S_{\!j} = Y_{k}$, skeleton points are compared with all skeletons on larger scales ($k+1$, \dots, $K$): points lying within a radius of one-fifth of the larger scale (an empirically optimized parameter) are removed. This procedure is repeated iteratively for all scales until the last pair ($Y_{K-1}, Y_{K}$) has been processed. The algorithm effectively removes duplicate skeleton points on smaller scales that overlap with skeletons on larger scales, while preserving the largest-scale skeletons. The resulting skeletons on scales $S_{\!j} = Y_{k}$ (hereafter denoted $Y_k$ for conciseness) uniquely trace the crests of their own filaments with matching widths $H_k$.

The scale-dependent skeletons are visualized in Figs.~\ref{fig:filament.skeleton.taurus} and \ref{fig:filament.skeleton.clouds.a}--\ref{fig:filament.skeleton.clouds.c}. For clarity, the skeletons are shown only in selected subregions of each cloud.

\subsection{Cutting skeletons into short segments}
\label{cut_segments}

Observed filamentary structures are complex -- they differ substantially from the idealized model of simple cylindrical structures with uniform physical properties. Indeed, filaments display large variations in their widths and crest values, and there is no physical reason for them not to have intrinsically different properties along their lengths. Nevertheless, a common approach has been to average profiles along entire skeletons \citep[e.g.,] []{Arzoumanian+2019}. Such an oversimplified description introduces biases due to the dissimilar, fluctuating profiles of different parts of long filaments \citep[e.g.,] []{Panopoulou+2017,Menshchikov+Zhang2026subm}. Furthermore, total filament lengths (and masses) cannot be estimated with certainty because unrelated fluctuations can easily fragment a long filament into disconnected pieces or artificially connect different filaments into seemingly continuous ones, as confirmed by our extensive experience with filament extractions in \textit{Herschel} images.

To improve our understanding of filament properties, profiles must be analyzed individually as a function of position along the skeleton. This can be achieved by dividing observed filaments into short segments and averaging profiles over each segment, following our previous work \citep{Zhang+2024}. This provides local measurements of profiles and derived physical properties, which is also important for comparisons with properties of sources that form at those locations along the filaments.

In this study, we divided the detected skeletons into 10-pixel (30{\arcsec}) segments, corresponding to physical lengths of 0.020 and 0.13\,pc at the nearest (Taurus, 140\,pc) and most distant (Vela\,C, 920\,pc) clouds, respectively. This constant angular length was adopted both for simplicity and to provide uniform averaging effects for the filament profiles across all clouds. As a preparatory step, we removed spurious small-scale skeletons with nearly circular shapes, arising as residuals from the source separation process in \textit{getsf}.

\subsection{Selecting acceptably good filament profiles}
\label{fil_select}

Filamentary structures are much more difficult to accurately analyze and measure than sources because their footprints occupy much larger image areas. Fluctuations in molecular clouds over long and wide filaments are large, making accurate reconstruction of extended filament backgrounds practically impossible. Furthermore, most filaments strongly overlap and blend with other structures (also with themselves, due to filament curvature) that are either physically nearby or simply in projection along the line of sight. No deblending has been applied to filaments in previous studies because no method currently exists for deblending filament profiles, making it unclear whether measured filament profiles describe the true properties of a single filament or those of a complex blend of several structures with unknown individual contributions.

The only way to overcome these serious problems is to carefully analyze the one-sided profiles of each extracted filament and select only those with regular shapes that are apparently unaffected by blending. Excluding heavily distorted or strongly fluctuating profiles improves the measurement accuracy of basic physical properties of filaments (their scale-dependent widths, slopes, outer radii, linear densities, etc.). We selected acceptably good profiles suitable for measuring widths and slopes using several criteria. For simplicity, the following description assumes that filament profiles are normalized to unity at their crests.

We required that good profiles in the \textit{getsf} catalogs must descend to at least the half-maximum level on both sides and applied additional selection criteria to each filament side independently. To exclude significant high-level anomalies, profile values must remain below 1.05 around the peak ($r < r_{0.7}$, the radius where the profile drops to 0.7 of the peak value). To exclude unreasonably wide spurious structures, filament profiles must not be too wide relative to the spatial scales on which they were detected ($H_k < 3 Y_k$). To mitigate the effects of blending with nearby structures, we identified a truncation radius $r_{\rm T}$ by locating local minima in the radial profiles beyond the peak; when no local minimum was found, the profile descends monotonically and we assigned $r_{\rm T}$ to the last radial point of the profile in the \textit{getsf} catalogs. We selected only profiles with high coefficients of determination ($\textsc{R}^2 > 0.97$) from the model fitting described in Sect.~\ref{fil_measure}.

Only those one-sided profiles that passed the above selection process were deemed acceptably good for measurements and further analysis. Other profiles were excluded to avoid inaccurate measurements or biases arising from various distortions, background fluctuations, or blending with other structures.

Figures~\ref{fig:filament.skeleton.taurus} and \ref{fig:filament.skeleton.clouds.a}--\ref{fig:filament.skeleton.clouds.c} display only those skeleton segments that have acceptably good one-sided filament profiles associated with them.

\subsection{Deriving physical properties from fitting filament profiles}
\label{fil_measure}

Previous studies analyzed observed profiles by fitting the Plummer function \citep[e.g., Eq.~(5) in] []{Arzoumanian+2019}, which does not account for the finite outer radius $R$ of filaments. This is because surface densities of finite structures do not resemble power laws toward their boundaries ($r \rightarrow R$), even if their volume density $\rho(r)$ follows a pure (truncated) power law there \citep[see] [for detailed discussion]{Menshchikov+Zhang2026subm}.

The full width at half maximum $H_k$ is a measure of filament widths in surface density images, defined analogously to source half-maximum sizes. The widths $H_k$ are estimated by \textit{getsf} from one-sided, background-subtracted surface density profiles $\hat{\Sigma}_{k}(r)$ at the half-maximum level of the crest values $\hat{\Sigma}_{{\rm C}k} \equiv \hat{\Sigma}_{k}(0)$, together with their slopes ${\rm d} \log \hat{\Sigma}_k(r) /{\rm d} \log r$ at each radial point $r$ and linear densities $\Lambda_k$. We recomputed all of these quantities by fitting an accurate surface density model to the filament profiles.

We applied the method developed by \cite{Menshchikov+Zhang2026subm} with a fitting function that accounts for the finite nature of surface density structures (see Appendix~\ref{app:fittingfun} for full derivation). The surface density profile (the fitting function) is given by
\begin{equation}
\Sigma(r) = \Sigma_{\rm C}\left(1+\left(2^{2/\gamma\!}-1\right)\left(\frac{2r}{w}\right)^2\right)^{-\gamma/2} \!\left(1-\left(\frac{r}{R}\right)^{\epsilon} \right)^{1/2}\,,
\label{surfdens_fittingfun}
\end{equation} 
where the square-root function describes the dependence of $\Sigma(r)$ on the geometry of the cylindrical boundary, $\Sigma_{\rm C}$ is the crest surface density, $\gamma$ is the intrinsic slope of surface density, $w$ is the intrinsic half-maximum width, $R$ is the boundary radius, and $\epsilon$ is an empirical exponent ensuring consistency with the volume density profile (Appendix~\ref{app:fittingfun}). The fitting function has only three independent parameters to be optimized: $\gamma$, $R$, and $\Sigma_{\rm C}$. All other parameters ($h$, $\beta$, $w$, and $\epsilon$) are derived from the analytical relations given in Appendix~\ref{app:fittingfun}. We adopted the fitting strategy described in \cite{Menshchikov+Zhang2026subm}, with fitting bounds of [0.01, 8] for the slope $\gamma$.

Although the filament component produced by \textit{getsf} on each spatial scale is background-subtracted (Fig.~\ref{fig:taurus_cut_profiles}), filament profiles are often blended with nearby structures or background fluctuations. Therefore, their endpoints at $r=r_{\rm T}$ frequently have substantial nonzero values, indicating the presence of residual backgrounds that must be removed for proper measurements. We estimated the background of each profile by linear interpolation between the profile values at $r=r_{\rm T}$ on both sides and subtracted the resulting background before fitting the model.

The volume density widths $h_k$ are provided directly by the fitting, while the surface density widths $H_k$ are obtained numerically from the fitted $\Sigma_{k}(r)$ profiles at their half-maximum levels. The axial volume density $\rho_{{\rm C}k} \equiv \rho_{k}(0)$ was computed by dividing $\Sigma_{{\rm C}k}$ by the numerically integrated dimensionless profile of Eq.~(\ref{volume_density}) along the radial direction ($-R_k \le r \le R_k$). The volume density slopes $\beta_k$ are derived from Eq.~(\ref{betaformula}) using the intrinsic surface density slopes $\gamma_k$, which are unaffected by the geometry factor in Eq.~(\ref{surfdens_fittingfun}). The contrasts $C_{\Sigma k}$ and $C_{\rho k}$, measuring the prominence of filament crests and axes above their respective background levels, are defined as
\begin{equation}
C_{\Sigma k} = \Sigma_{{\rm O}k}(0) / \Sigma_{{\rm O}k}(R), \quad
C_{\rho k} = \rho_{k}(0) / \rho_{k}(R),
\label{fil_contrasts}
\end{equation}
where $\Sigma_{{\rm O}k}(R)$ is the mean background level averaged over the profile endpoints in the original surface density image and $\rho_{k}(R)$ is the volume density at the profile edge. The linear densities $\Lambda_{\Sigma k}$ and $\Lambda_{\rho k}$ of filaments are computed by numerical integration of the respective profiles
\begin{equation}
\Lambda_{\Sigma k} = 2 \mu m_{\rm H} \int_{0}^{R} \!\Sigma_{k}(r)\,{\rm d}r, \quad
\Lambda_{\rho k} = 2 \mu m_{\rm H} \pi \int_{0}^{R} \!r \rho_{k}(r)\,{\rm d}r,
\label{fil_linear_densities}
\end{equation}
where $\mu = 2.8$ is the mean molecular weight per H$_2$ molecule and
$m_{\rm H}$ is the mass of the hydrogen atom. The two estimates of the linear densities should give identical results if the profiles in Eqs.~(\ref{surfdens_fittingfun}) and (\ref{volume_density}) are fully consistent; the actual relative differences over $\sim\!10^4$ filament segments have a median of 1\%, with values below 15\% in all cases, where the larger differences occur for shallow profiles with $\beta < 1$ for which the analytical approximations are less accurate.

The fitting of observed filament profiles $\hat{\Sigma}(r)$ with our model yielded high-quality results, with a median coefficient of determination $\textsc{R}^2 = 0.98$ and very small relative uncertainties in $\Sigma_{\rm C}$ and $R$ (median relative uncertainties of only $0.021$ and $0.031$, respectively). The relative uncertainties in the slopes $\gamma$ are larger (median $0.14$), indicating that $\gamma$ can be poorly constrained by the fitting and its individual values may carry substantial errors. To ensure reliability, all fitted profiles were required to satisfy the following quality criteria: $\textsc{R}^2 > 0.97$, a relative residual between the fit $\Sigma(r)$ and observed profile $\hat{\Sigma}(r)$, normalized by the mean value of $\hat{\Sigma}(r)$ over the fitting range, below 0.2, and -- for parameters $\beta$, $h$, $w$, and $\epsilon$ depending on $\gamma$ -- the diagonal element of the covariance matrix for $\gamma$ below a threshold of 2 \citep{Menshchikov+Zhang2026subm}.

\begin{figure*}[ht!]
\centering
\includegraphics[width=\hsize]{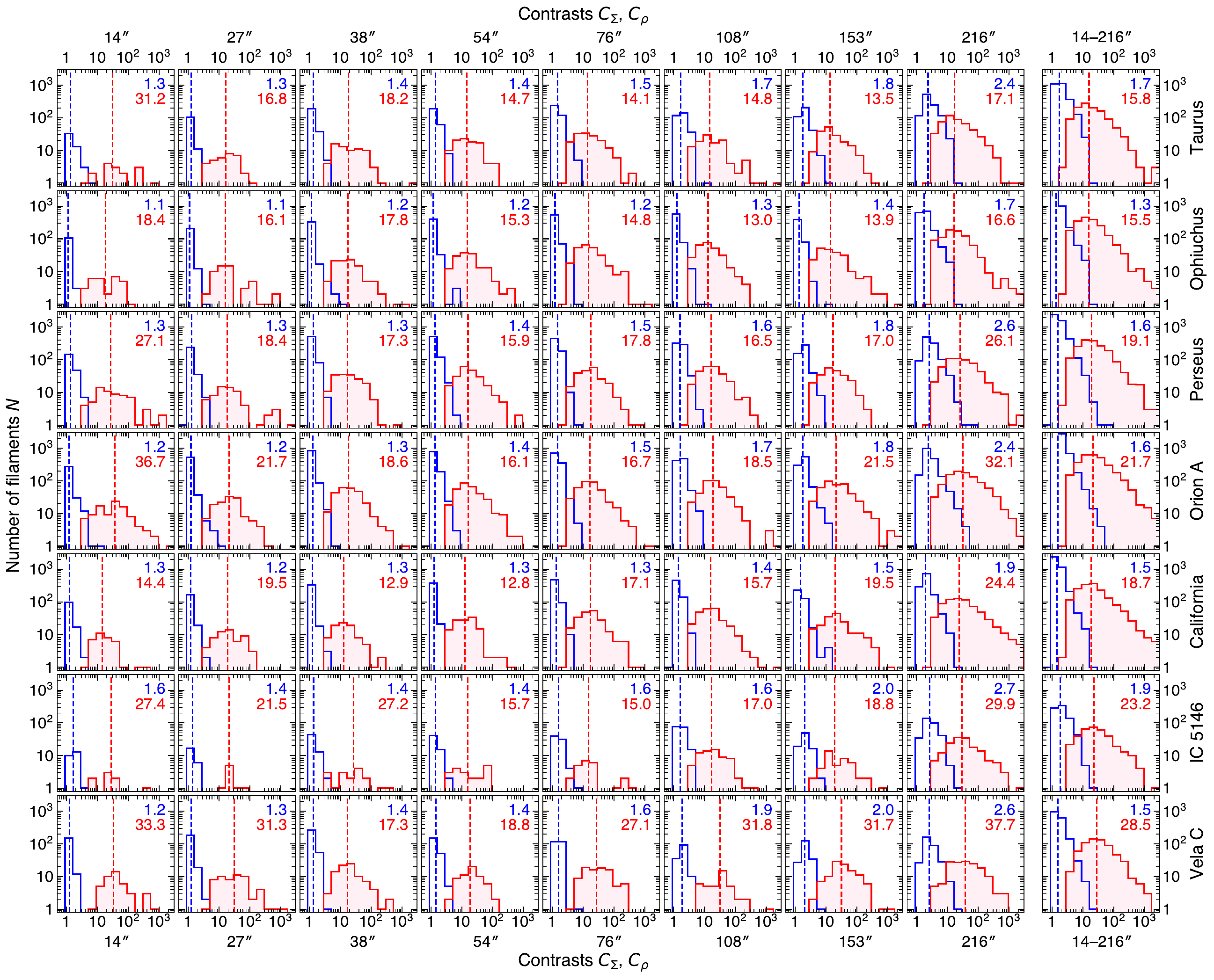}
\caption{
Scale-dependent distributions of surface and volume density contrasts $C_{\Sigma k}$ (blue) and $C_{\rho k}$ (red) of filaments in the molecular clouds, obtained from profiles $\Sigma(r)$ and $\rho(r)$, respectively (Sect.~\ref{fil_measure}). Dashed lines indicate median values $\tilde{C}_{\Sigma k}$ and $\tilde{C}_{\rho k}$, given in the panels as upper and lower values, respectively. Rightmost panels show distributions accumulated over all scales.
}
\label{fig:filament_contrasts}
\end{figure*}

\begin{figure*}[ht!]
\centering
\includegraphics[width=\hsize]{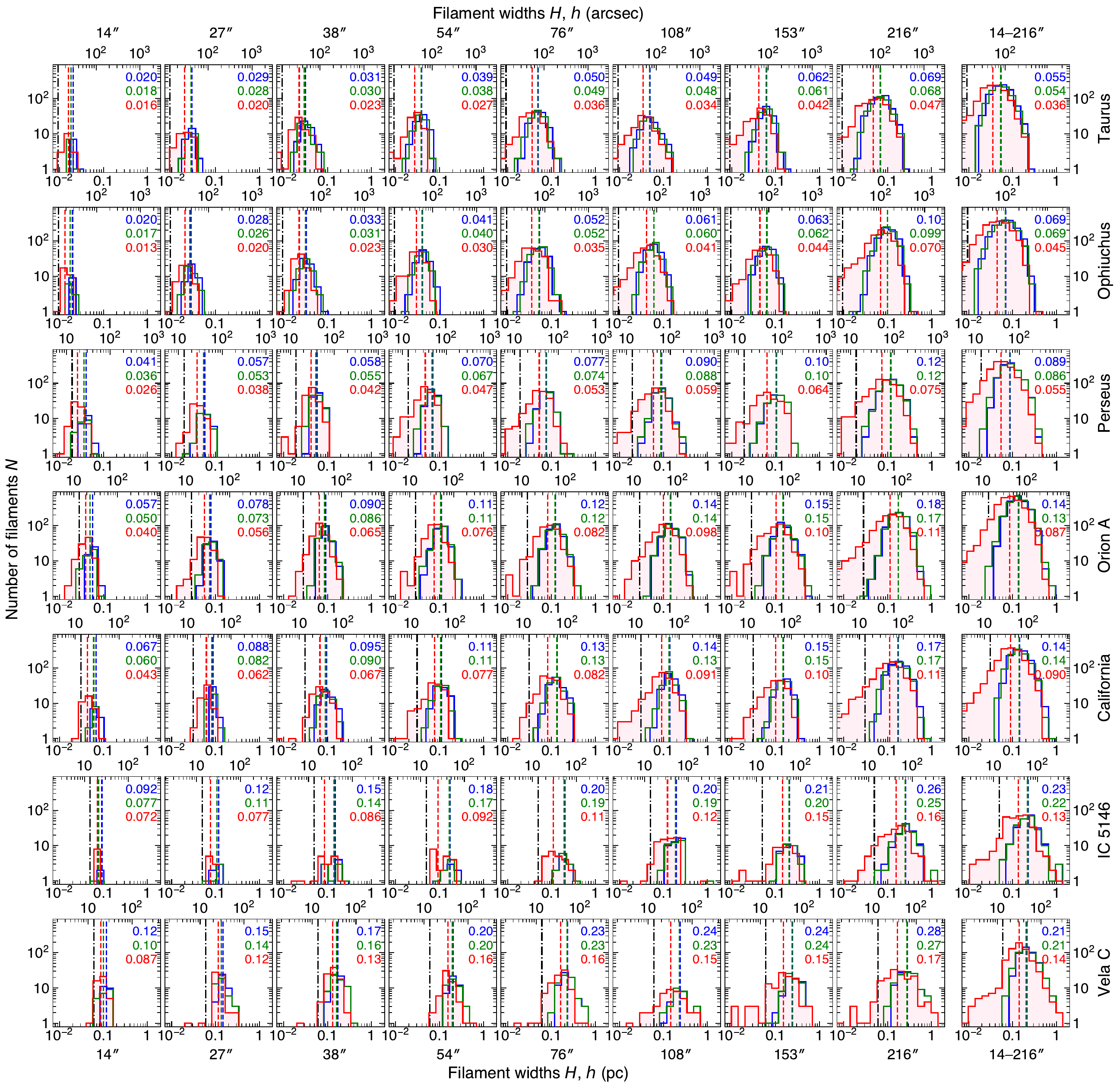}
\caption{
Scale-dependent distributions of the half-maximum widths $H_k$ (blue) and $h_{k}$ (red) of filaments in the molecular clouds, derived for the surface and volume density profiles $\Sigma(r)$ and $\rho(r)$, respectively (Sect.~\ref{fil_measure}). Also shown are the distributions of the deconvolved widths $\breve{H}_{k}$ (green), estimated from Eq.~(\ref{gauss_deconv}) for only those filaments that can be deconvolved with an accuracy better than 20\%. Dot-dashed lines indicate the angular resolution of $13.5${\arcsec} and dashed lines mark the median values $\tilde{H}_{k}$ and $\tilde{h}_{k}$ (given in the panels) of the distributions. The rightmost panels show the distributions accumulated over all scales.
}
\label{fig:WidthDistribution_shortavg}
\end{figure*}

\begin{figure*}[ht!]
\centering
\includegraphics[width=\hsize]{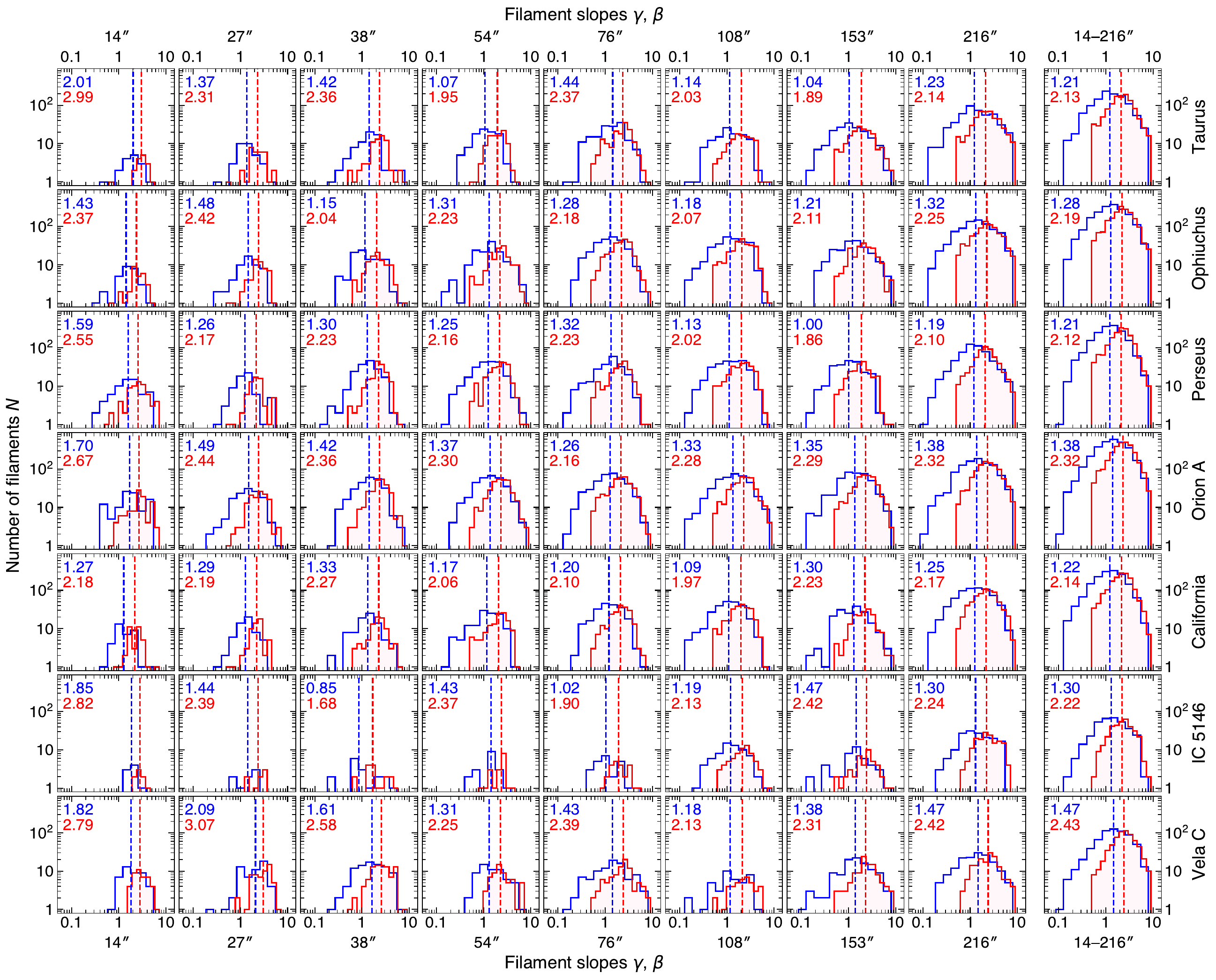}
\caption{
Scale-dependent distributions of surface and volume density slopes $\gamma_k$ (blue) and $\beta_k$ (red) of filaments in the molecular clouds, obtained from profiles $\Sigma(r)$ and $\rho(r)$, respectively (Sect.~\ref{fil_measure}). Dashed lines indicate median values $\tilde{\gamma}_k$ and $\tilde{\beta}_k$, given in the panels as upper and lower values, respectively. Only profiles passing the quality criteria (Sect.~\ref{fil_select}) are included. Rightmost panels show distributions accumulated over all scales.
}
\label{fig:slopeDistribution_shortavg}
\end{figure*}

\subsection{Convergence tests}
\label{convergence_tests}

To assess how angular resolution and distance affect the reliability of filament measurements, we performed a series of resolution convergence tests for each of the seven molecular clouds. We defined a set of additional angular resolutions $O_{i}$ of 27, 38, 54, 76, 108, and 216{\arcsec} ($i=1, 2, \dots, I$, with $I = 6$) and convolved the 13.5{\arcsec} resolution surface density images of each cloud to these resolutions, thereby simulating the effect of observing the same cloud at lower resolution or greater distance. For each $O_{i}$, we ran \textit{getsf} to separate the filament components using the maximum sizes $Y_k$ defined in our filament extractions (Sect.~\ref{maxsizes}), producing scale-dependent background-subtracted filament images on each spatial scale $S_{\!j} = Y_k$. Then we ran the \textit{getsf} utility \textit{fmeasure} to create catalogs of filament measurements for each $O_i$ using the existing scale-dependent skeletons from the original 13.5{\arcsec} resolution extractions (Sect.~\ref{trace_unique}), keeping the skeletons fixed so as to isolate the effect of resolution on profile measurements from its effect on skeleton tracing. Finally, we derived the scale-dependent filament widths $H_{ki}$, slopes $\gamma_{ki}$, and linear densities $\Lambda_{ki}$ in each molecular cloud for each resolution $O_i$, using the same approach as in the original extractions with $O = 13.5${\arcsec}. The results of these convergence tests are presented and discussed in Sect.~\ref{discuss_distance}.

\section{Results}
\label{results}

This section presents the results of our filament extraction analysis in seven molecular clouds: \object{Taurus}, \object{Ophiuchus}, \object{Perseus}, \object{Orion A}, \object{California}, \object{IC 5146}, and \object{Vela C}. The extracted filament skeletons are displayed in Figs.~\ref{fig:filament.skeleton.clouds.a}--\ref{fig:filament.skeleton.clouds.c} (Appendix~\ref{app:skeletons}). For profiles meeting our quality criteria, we compute median measurements along 10-pixel filament segments on each spatial scale $Y_{k}$. The resulting filament profiles analyzed below are illustrated in Fig.~\ref{fig:filament_radial_profiles_fitted_surface_volume} (Appendix~\ref{app:surfdensprofs}).

\subsection{Filament contrasts}
\label{fil_contrast}

Figure~\ref{fig:filament_contrasts} displays scale-dependent distributions of the contrasts $C_{\Sigma k}$ and $C_{\rho k}$ derived from the surface and volume density profiles, respectively (Eq.~(\ref{fil_contrasts})), along with the distributions accumulated over all scales. The contrast $C_{\Sigma k}$ is the most direct measure of filament prominence against the observed surface density background and is therefore most directly related to filament detectability, whereas $C_{\rho k}$ characterizes the prominence of the physical volume density distribution and depends on the fitted profile extent $\xi_{k} \equiv R_{k}/h_{k}$ and slope $\beta_{k}$.

The surface density contrast $C_{\Sigma k}$ increases systematically with spatial scale, with the median values $\tilde{C}_{\Sigma k}$ rising by factors of $\sim$\,1.5--2 from the range [1.1, 1.6] across clouds on the 14{\arcsec} scale to [1.7, 2.7] on the 216{\arcsec} scale. However, the volume density contrast $C_{\rho k}$ shows no clear systematic dependence on spatial scale, with only irregular fluctuations -- the median values $\tilde{C}_{\rho k}$ spanning the range [18, 37] across clouds on the 14{\arcsec} scale and [17, 38] on the 216{\arcsec} scale. The relationships between contrasts and scales can be approximated by power laws $\tilde{C}_{\Sigma} \propto Y^{0.19\pm 0.02}$ and
$\tilde{C}_{\rho} \propto Y^{-0.002\pm 0.04}$ for the combined sample (where the uncertainties are standard errors of the fits), with the exponents varying in the ranges [0.14, 0.27] and [$-$0.09, 0.13] among individual clouds.

The distributions of $C_{\Sigma k}$ are relatively narrow (largest values below $\sim$\,20), because the two-dimensional surface density is integrated across the entire embedding cloud, which raises the background level and limits the achievable contrast. The distributions of $C_{\rho k}$ are much broader, spanning three orders of magnitude, because this contrast measures the prominence of the filament axial volume density against the three-dimensional local background. The high values of $C_{\rho k}$ indicate that the distributions contain extended profiles ($\xi_{k} \gg 1$) and/or steep slopes $\beta_{k}$.

Broader distributions of $C_{\Sigma k}$ with increasing median values on larger scales indicate that wider filaments tend to have higher surface density contrasts. The broader distributions of $C_{\rho k}$ on larger scales suggest that wider filaments are also more prominent in their physical size and volume density relative to the local background.

\subsection{Filament widths}
\label{fil_widths}

Figure~\ref{fig:WidthDistribution_shortavg} displays the scale-dependent distributions of the half-maximum widths $H_{k}$ and $h_{k}$ for the median surface density profiles $\Sigma_k(r)$ and volume density profiles $\rho_k(r)$, respectively, along with distributions accumulated over all spatial scales.

For comparisons with previous observational studies of filaments, we also plotted the deconvolved widths $\breve{H}_{k}$ estimated using a simple Gaussian deconvolution formula
\begin{equation}
\breve{H}_{k}=H_{k} \left(1-\mathcal{R}_{k}^{-2}\right)^{1/2},
\label{gauss_deconv}
\end{equation}
where $\mathcal{R}_{k}\equiv H_{k}/O > 1$ is the profile resolvedness \citep{Menshchikov2023}. Following our demonstration that Eq.~(\ref{gauss_deconv}) is accurate to within 20\% only when $\mathcal{R}\ga 1+7\beta^{-2}$ \citep[Appendix~A.3 in] []{Menshchikov+Zhang2026subm}, we show $\breve{H}_{k}$ distributions only for profiles satisfying this condition. This condition is only an approximate indicator of the applicability domain of Eq.~(\ref{gauss_deconv}), because the derived slope $\beta$ may have large systematic errors. We emphasize, however, that the deconvolved surface density widths $\breve{H}$ cannot always be interpreted as the physical volume density widths, because $h \ll H$ for shallower profiles with slopes $\beta \la 2$ \citep{Menshchikov+Zhang2026subm}.

As expected, filaments detected on smaller scales exhibit narrower widths, while those on larger scales are progressively wider across all molecular clouds. The median surface density widths $\tilde{H}_k$ range from [0.020, 0.12]\,pc on the 14{\arcsec} scale to [0.069, 0.28]\,pc on the 216{\arcsec} scale. Similarly, the median volume density widths $\tilde{h}_k$ span [0.013, 0.087]\,pc and [0.047, 0.17]\,pc on these scales, respectively. Distributions of the deconvolved widths $\breve{H}_{k}$ on large scales $Y_{k}\ga 50${\arcsec} become very similar to those of the measured widths $H_{k}$, which reflects the sub-selection of only those profiles with sufficiently large resolvedness; applying deconvolution to all profiles would likely produce broader distributions shifted toward smaller values. Distributions of the volume density widths $h_{k}$ are broader and shifted toward smaller values by factors of $\sim$\,2--5, indicating the presence of shallower profiles with $\beta_{k} \la 2$.

The scale-dependent behavior of filament widths is expected, as imaged structures are generally most detectable on matching spatial scales \citep[$H_k \approx Y_k$; Appendix B in] []{Menshchikov2021method}. The relationships between widths and scales follow power laws $\tilde{H} \propto Y^{0.50\pm 0.09}$ and $\tilde{h} \propto Y^{0.37\pm 0.09}$ for the combined sample, with the exponents varying in the ranges [0.41, 0.65] and [0.21, 0.56] among individual clouds. At the angular resolution of $O = 13.5${\arcsec}, filaments appear resolved (median $\tilde{\mathcal{R}}_k > 2$) even on the smallest scale of 14{\arcsec} and become increasingly better resolved on larger scales. However, \cite{Menshchikov+Zhang2026subm} demonstrated that filaments with shallow density slopes remain effectively unresolved even when $\mathcal{R}_{k}\gg 1$.

Filaments detected on larger scales exhibit broader and less peaked $H_{k}$ and $h_{k}$ distributions and appear more abundant than smaller-scale filaments. Since our analysis focuses on short filament segments, this apparent increase in abundance suggests that the extracted filaments are physically longer on larger spatial scales, since longer filaments produce more 10-pixel segments. Indeed, we find that the median length of filaments increases with scale as a power law $\tilde{L} \propto Y^{0.70 \pm 0.11}$ for the combined sample, with the exponent varying in the range [0.64, 0.77] among individual clouds. In contrast, the total number of extracted filaments shows no clear scale dependence.

\begin{figure*}[ht!]
\centering
\includegraphics[width=\hsize]{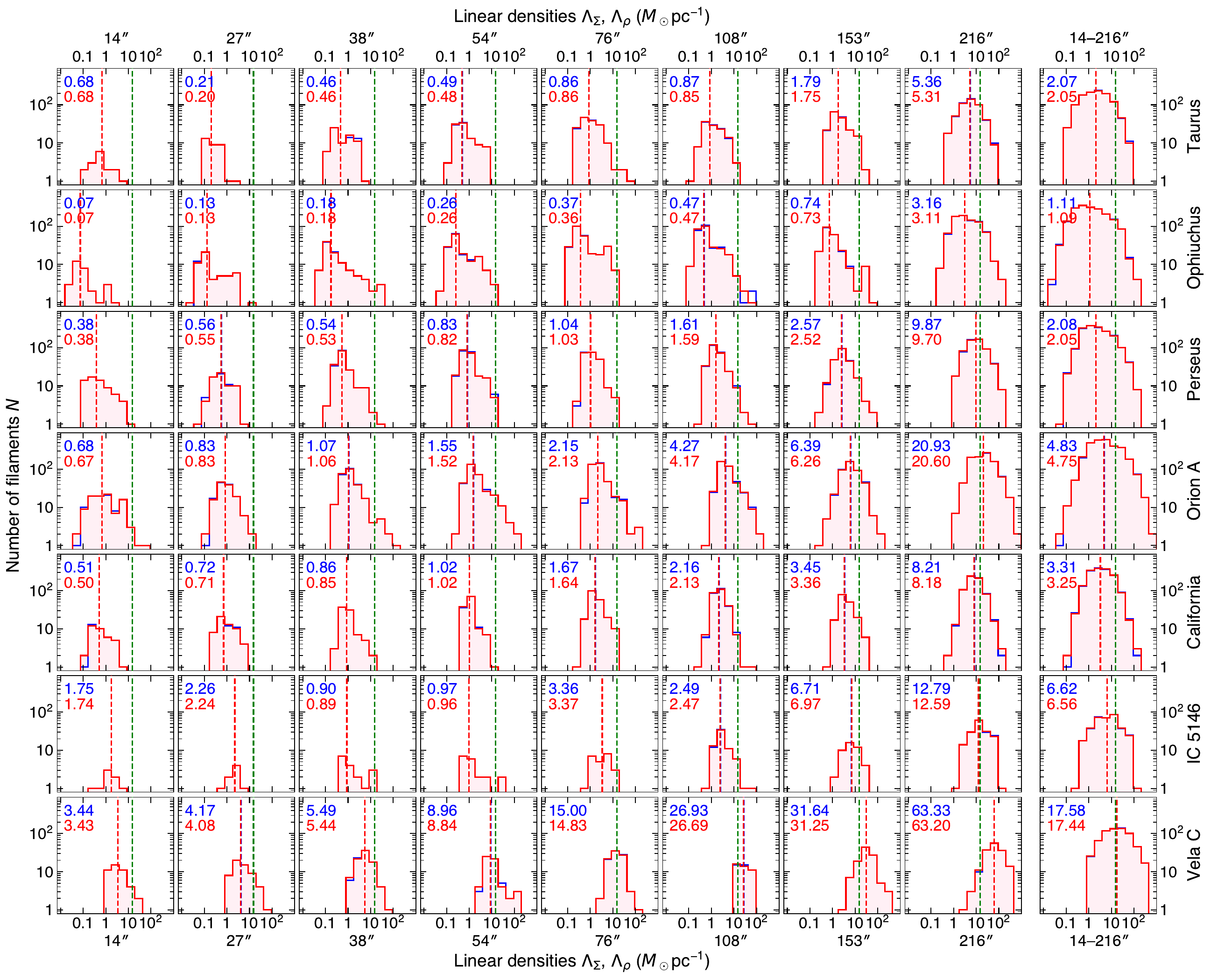}
\caption{Scale-dependent distributions of linear densities $\Lambda_{\Sigma k}$ (blue) and $\Lambda_{\rho k}$ (red) of filaments in the molecular clouds, integrated from profiles $\Sigma(r)$ and $\rho(r)$, respectively (Sect.~\ref{fil_measure}). The two linear density measures are practically identical (within 3\%). Dashed lines in the seven upper rows indicate median values $\tilde{\Lambda}_{\Sigma k}$ and $\tilde{\Lambda}_{\rho k}$, listed as the upper and lower values, respectively. Green dashed lines indicate the critical value of 15\,$M_\odot$\,pc$^{-1}$ (see Sect.~\ref{discuss_stability}) for reference. Rightmost panels show distributions accumulated over all scales. 
}
\label{fig:lindensDistribution}
\end{figure*}

\subsection{Filament slopes}
\label{fil_slopes}

Figure~\ref{fig:slopeDistribution_shortavg} shows scale-dependent distributions of the slopes $\gamma_{k}$ and $\beta_{k}$ (Sect.~\ref{fil_measure}) of the surface and volume density profiles, along with the distributions accumulated over all scales.

Across all molecular clouds, filaments detected on the smallest scales $Y_{k}\la 30${\arcsec} appear to have somewhat steeper profiles on average than those detected on larger scales. The median intrinsic surface density slopes $\tilde{\gamma}_k$ span [1.3, 2.0] on the 14{\arcsec} scale, decreasing to [1.2, 1.5] on the 216{\arcsec} scale. The median volume density slopes $\tilde{\beta}_k$ span [2.2, 3.0] and [2.1, 2.4] on these scales, respectively. We find that the median slopes decrease with scale as power laws $\tilde{\gamma} \propto Y^{-0.092 \pm 0.02}$ and $\tilde{\beta} \propto Y^{-0.062 \pm 0.02}$ for the combined sample, with the exponents varying in the ranges [$-$0.18, $-$0.02] and [$-$0.12, $-$0.011] among individual clouds. The physical significance of this trend is, however, difficult to assess given the limited statistics and large uncertainties in background subtraction.

Filaments detected on specific spatial scales exhibit fairly broad $\gamma_{k}$ and $\beta_{k}$ distributions. When accumulated over all scales (14--216{\arcsec}), the distributions become somewhat wider than the individual scale-dependent distributions, reflecting the spread in median values across scales. The accumulated distributions have $\tilde{\gamma} \approx$\,1.2--1.5 and $\tilde{\beta} \approx$\,2.1--2.4, with little variation among clouds and values almost identical to those on the largest scale.

The distributions on the largest 216{\arcsec} scale have their upper bins at $\gamma_{k} = 8$ and $\beta_{k} = 9$ (since $\beta \approx \gamma + 1$ for large values), reflecting the upper bound $\gamma_{\rm max} = 8$ adopted in the profile fitting. A substantial fraction ($\sim$\,40\%) of the profiles satisfying the fitting quality criteria (Sect.~\ref{fil_measure}) are found at this boundary, indicating that the fitting is highly sensitive to $\gamma$ and tends to produce unreliably large values, particularly for profiles with low resolvedness. Following the results of the model study by \cite{Menshchikov+Zhang2026subm}, we therefore exclude profiles with $\gamma_{k} = 8$ to improve the reliability of the fitting results.

\begin{figure*}[ht!]
\centering
\includegraphics[width=\hsize]{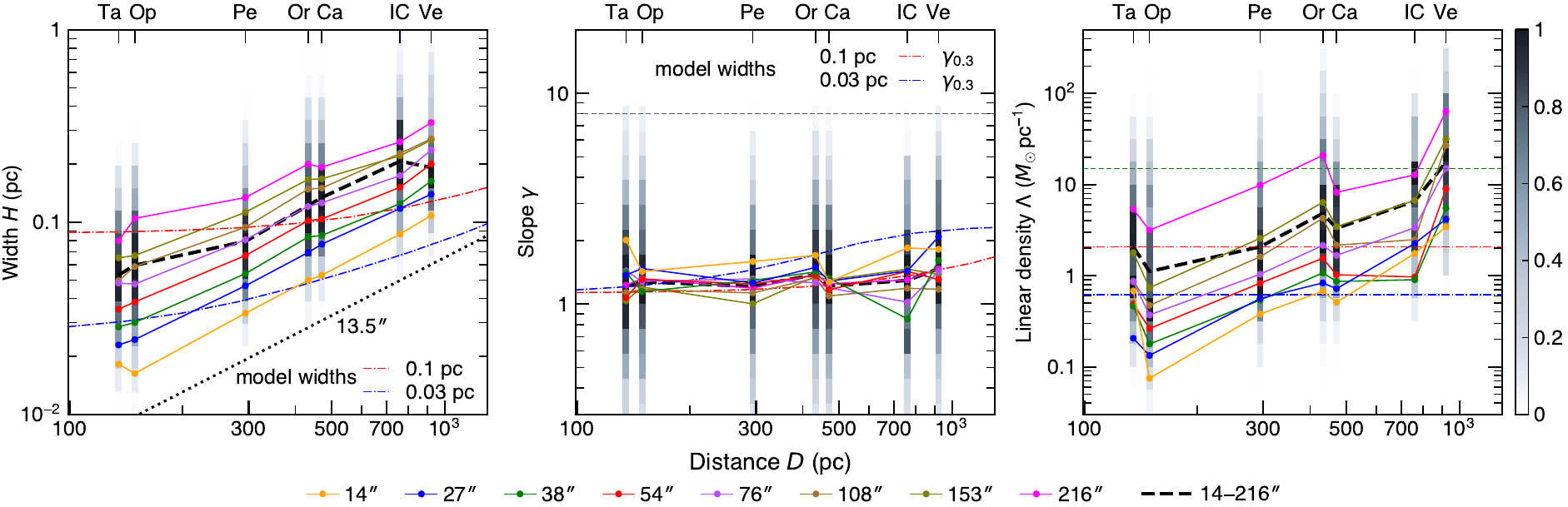}
\caption{
Median half-maximum widths $H_k$, profile slopes $\gamma_k$, and linear densities $\Lambda_k$ of scale-dependent filaments in the molecular clouds. Two-letter name abbreviations are shown above the upper axes at the cloud distances. To avoid overlaps, results for \object{Taurus} and \object{Ophiuchus} are shifted by $\pm$5\,pc, respectively. The plotted quantities are computed from profiles meeting the quality criteria for filament segments on spatial scales indicated in the middle panel. Lines connecting values for individual clouds are shown for visual guidance only. Grayscale bars show normalized distributions of $H$, $\gamma$, and $\Lambda$ over all scales (14--216{\arcsec}, Figs.~\ref{fig:WidthDistribution_shortavg}--\ref{fig:lindensDistribution}), with thick dashed lines linking their median values. Thin dot-dashed curves show quantities for two isolated filament models with true $H$ of 0.1 and 0.03\,pc, boundary radius $R=2H$, and intrinsic $\gamma = 1$ (Sect.~\ref{fil_distances}). The apparent model slope $\gamma_{0.3}$ was evaluated at 30\% of the filament crest value. For the model $\Lambda_k$ (distance-independent), the 0.1\,pc curve is arbitrarily fitted to the thick dashed line at $D = 140$\,pc. For reference, the dotted line in the left panel shows the 13.5{\arcsec} resolution and the dashed horizontal lines in the middle and right panels indicate the fitting bound $\gamma_{\rm max\!} = 8$ and critical value $\Lambda_{\rm c} = 15$\,$M_\odot$\,pc$^{-1}$ (see Sect.~\ref{discuss_stability}), respectively.
}
\label{fig:width_slope_linedens_dist}
\end{figure*}

\subsection{Filament linear densities}
\label{fil_lindens}

Distributions of linear densities may be referred to as the filament linear density function (FLDF), in analogy with the core mass function (CMF) for dense cores \citep[e.g., Sect.~3.3.3 in] []{Zhang+2024}, and their properties are important for a deeper understanding of star formation in filamentary structures.

Figure~\ref{fig:lindensDistribution} presents scale-dependent distributions of the linear densities $\Lambda_{\Sigma k}$ and $\Lambda_{\rho k}$ derived from the surface and volume density profiles, respectively (Eq.~(\ref{fil_linear_densities})), along with the distributions accumulated over all scales. Both estimates of linear density are consistent and have practically identical distributions; we therefore refer to them collectively as $\Lambda_{k}$. Figure~\ref{fig:lindensDistribution} also shows the differential and cumulative distributions ($N$, $N_{\rm C}$) combined over all molecular clouds for improved statistical significance and, for comparison, the distributions derived when averaging filament profiles along entire filament lengths.

Across all molecular clouds, filaments detected on larger scales have progressively higher linear densities. The median values $\tilde{\Lambda}_{k}$ in the range [0.07, 3.4]\,$M_{\odot}$\,pc$^{-1}$ on the 14{\arcsec} scale increase to [3.2, 63]\,$M_{\odot}$\,pc$^{-1}$ on the 216{\arcsec} scale. Our results demonstrate that the relationship between the median linear densities and spatial scales follows a power law $\tilde{\Lambda} \propto Y^{1.01 \pm 0.18}$, with the exponent varying in the range [0.72, 1.2] among individual clouds. The statistical properties of the combined FLDF across all seven clouds will be presented in a subsequent paper.

\subsection{Dependence on distance}
\label{fil_distances}

Figure~\ref{fig:width_slope_linedens_dist} presents filament widths, slopes, and linear densities for the seven molecular clouds against their distances from 140 to 920\,pc, where, in addition to the median scale-dependent values accumulated over spatial scales of 14--216{\arcsec}, grayscale bars show the full distributions displayed in the rightmost panels of Figs.~\ref{fig:WidthDistribution_shortavg}--\ref{fig:lindensDistribution}.

To better understand the results, we explored simple models of an isolated straight filament without any background. The filament has a Gaussian core $G(r)$ and power-law wings $P(r) \propto r^{-\gamma_0}$, smoothly joined at the radial point where ${\rm d}G(r)/{\rm d}r = -\gamma_0$, ensuring a smooth transition between the two components. We created two models with widths $H_0$ of 147.3 and 44.2{\arcsec}, corresponding to linear widths of 0.1 and 0.03\,pc at a distance of 140\,pc, slope $\gamma_0 = 1$, and the same crest surface density $\Sigma_{\rm C}$. The model filaments were then convolved with a set of Gaussian beams in the range from 13.5 to 2443{\arcsec}. In effect, the convolved models simulated observations of the filaments with resolution $O = 13.5${\arcsec} from distances $D$ between 140\,pc and 25.3\,kpc, covering the full range from fully resolved to completely unresolved filaments; the upper distance limit is chosen purely to sample the unresolved regime and does not represent realistic observations. The filament resolvedness $\mathcal{R}$ in the ranges of 15.5--1.06 and 4.7--1.03 for the wider and narrower filament, respectively, confirms that this range is well sampled.

Measured half-maximum widths $H$, slopes $\gamma$, and linear densities $\Lambda$ from the simulated observations of the model filaments are shown in Fig.~\ref{fig:width_slope_linedens_dist} as functions of their distance $D$. As expected, the observed model widths $H(D)$ increase as the initially resolved filament becomes less resolved and then completely unresolved at progressively larger distances, asymptotically approaching the beam size line $O(D)$.

However, the observed model slope $\gamma(r|l,D)$ depends also on radius $r$ or level $l$. For illustration, we show only the slope $\gamma_{0.3}(D)$, where $l_{0.3} \equiv 0.3\,\Sigma_{\rm C}$. The slope changes only slightly when the Gaussian core is well resolved, whereas it becomes progressively steeper, approaching the shape of the Gaussian beam, when both the Gaussian core and power-law wings become progressively diluted within the beam toward completely unresolved filaments.

In contrast, linear densities of isolated, background-free filaments do not depend on distance (angular resolution), since filament mass and length are unaffected by convolution. Therefore, $\Lambda(D)$ is constant and, given the same $\Sigma_{\rm C}$ for both models, the narrower filament has lower linear density. The comparison of these model predictions with the observational data in Fig.~\ref{fig:width_slope_linedens_dist} is discussed in Sect.~\ref{discuss_distance}.

\subsection{Correlations among filament properties}
\label{sec:correlations}

Interrelations among filament parameters are presented in Appendix~\ref{app:scatterplots} for the combined dataset across all clouds and scales.

Figure~\ref{fig:combined_width_relations} demonstrates that key relations established by our finite-cylinder model (Appendix~\ref{app:fittingfun}) are confirmed by the data. The surface density width $H$ scales almost linearly with the intrinsic width $w$ ($H \propto w^{0.80}$), whereas $h$ increases more slowly ($h \propto w^{0.59}$). The volume density slope $\beta$ decreases with increasing profile extent $\xi \equiv R/h$ ($\beta \propto \xi^{-0.37}$), and $\gamma$ increases with $R/w$ ratio ($\gamma \propto (R/w)^{0.72}$). Because $\beta$ is derived from $\gamma$ via the model (Eq.~(\ref{betaformula})), the two quantities are not independent; the observed tight correlation $\gamma \propto \beta^{1.37}$ ($r_{\rm P}=0.99$) therefore reflects the internal consistency of our fitting procedure rather than an independent observational result. Nevertheless, the derived $\beta$ values are systematically larger than $\gamma$, as expected for filaments with finite boundary radii $R$. The width ratios $w/H$ and $h/H$ vary systematically with $\beta$ ($w/H \propto \beta^{-0.44}$, $h/H \propto \beta^{0.63}$). Both ratios approach unity for steep profiles ($\beta \ga 4$) and deviate strongly for shallow profiles ($\beta \la 2$), demonstrating that surface density widths overestimate the true physical extent of filaments with shallow volume density profiles, in full agreement with the empirical relations derived in \cite{Menshchikov+Zhang2026subm}.

Figure~\ref{fig:combined_filament_properties} shows that the surface density width $H$ is weakly correlated with crest surface density ($H \propto \Sigma_{\mathrm{C}}^{0.23}$, Pearson correlation coefficient $r_{\rm P}=0.39$), while the volume density width $h$ is weakly anticorrelated with axial volume density ($h \propto \rho_{\mathrm{C}}^{-0.20}$, $r_{\rm P}=-0.30$). The linear density $\Lambda_{\Sigma}$ exhibits a strong positive correlation with crest surface density ($\Lambda_{\Sigma} \propto \Sigma_{\mathrm{C}}^{1.19}$, $r_{\rm P}=0.91$), consistent with the relation $\Lambda_{\Sigma} \propto \Sigma_{\mathrm{C}} H$ derived from the width correlation. Similarly, $\Lambda_{\rho}$ is positively correlated with axial volume density ($\Lambda_{\rho} \propto \rho_{\mathrm{C}}^{0.77}$, $r_{\rm P}=0.59$), confirming that linear density -- a key parameter for gravitational stability -- is linked to the axial volume densities of filaments.

Figure~\ref{fig:all_properties_vs_resolvedness} reveals that the surface density slopes $\gamma$ decrease weakly with resolvedness ($\gamma \propto \mathcal{R}^{-0.14}$), while the intrinsic widths $w$ and contrasts $C_{\Sigma}$ increase, as expected. The volume density contrasts tend to decrease as $C_{\rho} \propto \mathcal{R}^{-0.67}$, indicating that better-resolved filaments have relatively lower axial prominence against the three-dimensional background. This trend is consistent with flatter slopes $\beta$ and/or decreasing extents $\xi$ of the volume density profiles. The outer-radius ratios $R/w$, $R/H$, and $R/h$ all decline with $\mathcal{R}$, showing that the profile extent decreases for filaments with higher resolvedness.

\section{Discussion}
\label{discuss}

We first note an important methodological difference between this work and most previous observational studies of filaments. Filament properties in this work are measured for short 10-pixel segments along filament crests, rather than averaged over entire filament lengths as in most previous studies. This segment-based approach captures local variations in width, slope, linear density, and contrast along filament crests, providing a more detailed picture of filament properties than global averages. As a consequence, the distributions of measured properties presented here are broader than those from whole-filament studies, reflecting genuine physical variations along filament lengths, though short segments may also contribute additional measurement scatter compared to averaged profiles. Furthermore, local measurements are physically more appropriate for studying filament fragmentation into dense cores, since the conditions for gravitational instability are determined by the local linear density and profile shape rather than by properties averaged over the entire filament length. Direct comparisons with results from previous studies based on whole-filament averages should therefore be interpreted with this difference in mind.

\subsection{Volume density profiles: new observational constraints}
\label{discuss_vol_dens}

The derivation of volume density profiles $\rho(r)$ from the observed surface density profiles $\Sigma(r)$ is a key methodological contribution of this work. Previous observational studies of filaments have been largely limited to surface density profiles, leaving the three-dimensional structure of filaments essentially unconstrained. Our results provide direct observational estimates of volume density profiles for a large sample of filaments across seven molecular clouds.

The median volume density slopes $\tilde{\beta} \approx 2.1$--$2.4$ found across all clouds and spatial scales fall well below the value $\beta = 4$ expected for an isothermal cylinder in hydrostatic equilibrium \citep{Ostriker1964}, suggesting that the observed filaments are generally not well described by this idealized model. Values of $\beta \la 2$ found in a significant fraction of profiles indicate shallow density distributions that are inconsistent with gravitationally bound, pressure-supported cylinders. These shallow profiles may represent filaments that are either unbound, externally compressed, or still in the process of formation through large-scale converging flows.

A key result of our analysis is the systematic difference between the surface and volume density widths, $H$ and $h$, for profiles with shallow slopes. For $\beta \la 2$, the half-maximum width of the volume density profile $h$ is significantly smaller than the corresponding surface density width $H$, with distributions of $h_k$ shifted by a factor of $\sim 2$ toward smaller values relative to $H_k$. This demonstrates observationally that surface density widths overestimate the true physical width of filaments with shallow profiles. Consequently, measurements of filament widths based solely on surface density profiles -- as in all previous observational studies -- may systematically overestimate the physical extent of a significant fraction of filaments.

The volume density contrasts $C_{\rho}$ are dramatically higher than the surface density contrasts $C_{\Sigma}$, with median values of $\tilde{C}_{\rho} \approx 17$--$52$ compared to $\tilde{C}_{\Sigma} \approx 1.1$--$2.7$ across all scales and clouds. This large difference reflects the fundamental distinction between the two quantities: the surface density contrast is diluted by integration along the line of sight through the entire embedding cloud, whereas the volume density contrast measures the local three-dimensional prominence of the filament against its immediate background. The high values of $C_{\rho}$ confirm that filaments are physically far more prominent structures within their parent clouds than their projected appearance suggests, corresponding to volume density enhancements of one to two orders of magnitude above the local background. This has direct implications for assessments of filament stability and star formation potential, since gravitational instability is governed by the local linear density and volume density distribution rather than the projected surface density alone.

\begin{figure*}[ht!]
\centering
\includegraphics[width=0.97\hsize]{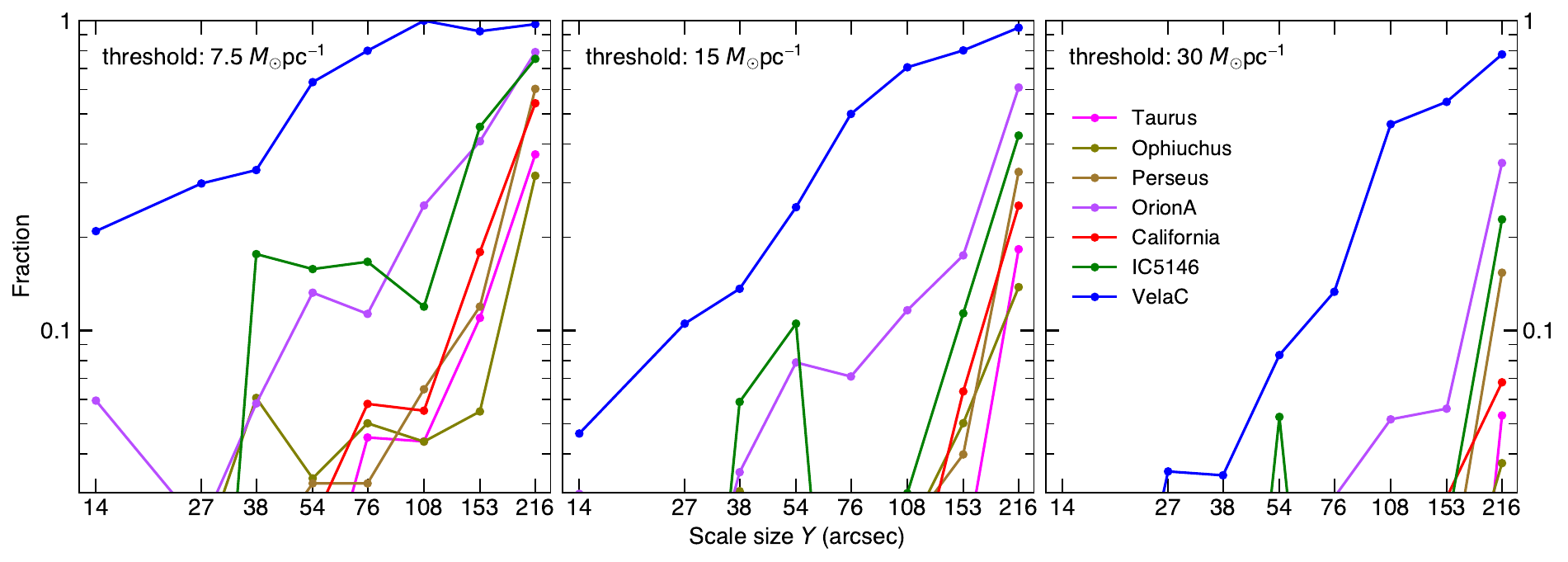}
\caption{
Fractions of filament segments with supercritical linear densities as a function of the spatial scale. The three thresholds $\Lambda_{\mathrm{c}}/2$, $\Lambda_{\mathrm{c}}$, and $2\Lambda_{\mathrm{c}}$ are indicated in the panels. Colored lines represent the seven molecular clouds studied in this work. Fractions are computed for filament segments that meet the quality criteria (Sect.~\ref{fil_measure}) and have well-determined $\gamma$ values.
}
\label{fig:supercritical_fractions}
\end{figure*}

\begin{figure*}
\centering
\includegraphics[width=1.0\hsize]{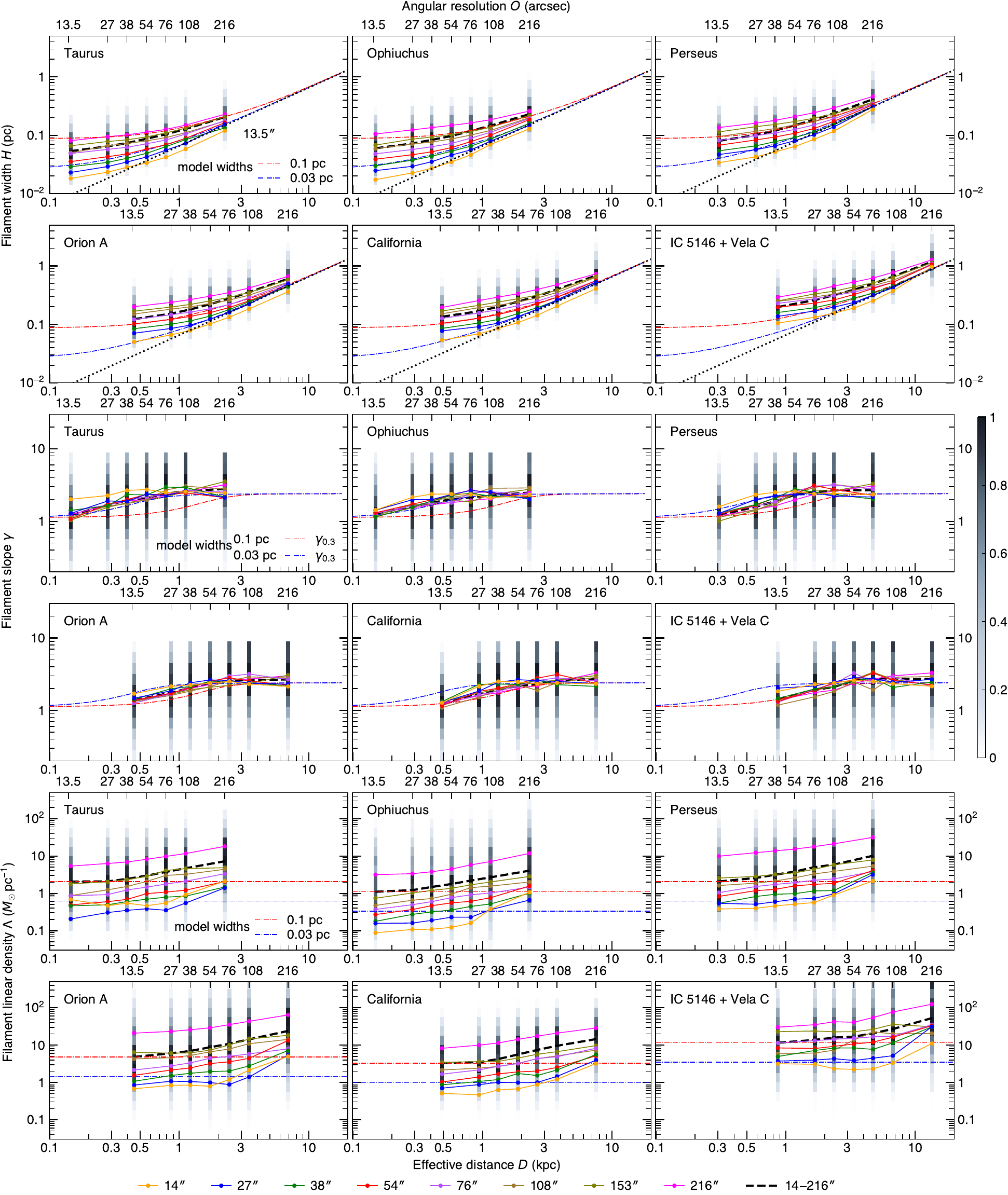}
\caption{
Convergence tests for the molecular clouds. Plots show measured half-maximum widths $H_k$, profile slopes $\gamma_k$, and linear densities $\Lambda_k$ of filaments as a function of cloud distance $D$ (lower axes) or angular resolution $O$ (upper axes). Colored lines connect median values at resolutions of 13.5, 27, 38, 54, 76, 108, and 216{\arcsec}, with thick dashed lines representing medians over all scales (14--216{\arcsec}). Gray-scale bars show normalized distributions of $H_k$, $\gamma_k$, and $\Lambda_k$ for all scales (Figs.~\ref{fig:WidthDistribution_shortavg}--\ref{fig:lindensDistribution}). Thin dot-dashed curves show quantities for two isolated filament models with true $H$ of 0.1 and 0.03\,pc, boundary radius $R=2H$ (a substantially extended model), and intrinsic $\gamma = 1$ (Sect.~\ref{fil_distances}). The apparent model slope $\gamma_{0.3}$ was evaluated at 30\% of the filament crest value. For the model $\Lambda_k$ (distance-independent), the 0.1\,pc curves are normalized to match the thick dashed lines at 13.5{\arcsec} resolution. Results for \object{IC\,5146} and \object{Vela\,C} are combined into a co-added sample at an average distance of 840\,pc to increase sample size.
}
\label{fig:width_slope_linearden_vs_distance}
\end{figure*}

\subsection{Scale dependence of filament properties}
\label{discuss_scale_depend}

The filament widths $H$ and $h$, slopes $\gamma$ and $\beta$, and linear densities $\Lambda$ all show clear dependencies on the spatial scales $Y$ on which the filaments are detected. The power-law relations $\tilde{H} \propto Y^{0.50}$, $\tilde{h} \propto Y^{0.37}$, and $\tilde{\Lambda} \propto Y^{1.01}$ hold across all seven molecular clouds, with relatively small cloud-to-cloud variations in the exponents. The slopes $\tilde{\gamma}$ and $\tilde{\beta}$ show a weaker but systematic decrease with scale, following $\tilde{\gamma} \propto Y^{-0.092}$.

The scale dependence of filament widths is a natural consequence of the multiscale analysis: structures are most detectable on spatial scales matching their characteristic size \citep[$H \approx Y$; Appendix B in] []{Menshchikov2021method}. However, the specific form of the width--scale relation carries physical information. The exponents of $\sim$\,0.37--0.50 for $\tilde{H}$ and $\tilde{h}$ are consistent with a picture in which filament widths reflect the characteristic size of the structures from which they are assembled. The linear scaling $\tilde{\Lambda} \propto Y^{1.01}$ implies that the linear density of filaments scales almost directly proportionally with spatial scale, meaning that larger-scale filaments are substantially more massive per unit length, which has direct consequences for their gravitational stability discussed in Sect.~\ref{discuss_stability}.

The weak but systematic decrease of slopes with scale ($\tilde{\gamma} \propto Y^{-0.092}$) indicates that filaments detected on larger scales tend to have somewhat shallower profiles on average. This is consistent with the expectation that larger-scale structures encompass more diffuse material in their outer regions, reducing the effective profile slope. However, given the limited dynamic range of the exponent variation ($[-0.18, -0.02]$ among individual clouds) and the uncertainties in background subtraction, this trend should be regarded with caution.

The relatively small cloud-to-cloud variations in the power-law exponents suggest that the scale dependence of filament properties is a robust feature of molecular cloud structure, largely independent of the particular physical conditions prevailing in individual clouds. The surface density contrasts $C_{\Sigma k}$ also follow a power law $\tilde{C}_{\Sigma} \propto Y^{0.19}$, indicating that filaments become progressively more prominent against their background on larger spatial scales. In contrast, the volume density contrasts $C_{\rho k}$ show no systematic scale dependence, suggesting that the intrinsic three-dimensional prominence of filaments is determined by local physical conditions rather than by the spatial scale of detection.

\subsection{Gravitational stability of filaments}
\label{discuss_stability}

The critical linear density $\Lambda_{\rm c} = 2c_{\rm s}^2/G$, where $c_{\rm s}$ is the isothermal sound speed and $G$ is the gravitational constant, of an isothermal cylinder in hydrostatic equilibrium \citep{Ostriker1964} is commonly used as a criterion to assess whether filaments are gravitationally unstable and therefore capable of fragmenting into dense cores \citep{Zhang+2020}. For a typical dust temperature of 10\,K, $\Lambda_{\rm c} \approx 15\,M_\odot$\,pc$^{-1}$, assuming thermal coupling between dust and gas in the dense interiors of molecular filaments, with the standard value $c_{\rm s} \approx 0.2$\,km\,s$^{-1}$. However, as discussed in \cite{Li+2023}, the applicability of this criterion to observed filaments is uncertain, because the assumptions of the idealized model -- infinite length, isothermality, hydrostatic equilibrium, negligible magnetic fields, and a uniform background -- are generally not satisfied by the complex filaments observed in molecular clouds. The critical linear density should therefore be regarded only as an order-of-magnitude indicator of gravitational instability. We adopt an uncertainty of a factor of 2 in $\Lambda_{\rm c}$ as a rough but reasonable choice, noting that \cite{Li+2023} used a factor of 3, which we consider too conservative for our more diverse sample of clouds. This defines three regimes: filaments with $\Lambda < \Lambda_{\rm c}/2 \approx 7.5\,M_\odot$\,pc$^{-1}$ as definitively subcritical, those with $\Lambda > 2\Lambda_{\rm c} \approx 30\,M_\odot$\,pc$^{-1}$ as definitively supercritical, and those in between as having ambiguous gravitational state, though the appropriate uncertainty factor may differ among clouds depending on their physical conditions \citep{Arzoumanian+2019}.

Figure~\ref{fig:supercritical_fractions} shows the fractions of filament segments exceeding the three thresholds $\Lambda_{\rm c}/2$, $\Lambda_{\rm c}$, and $2\Lambda_{\rm c}$ as functions of spatial scale for each of the seven molecular clouds. The fractions increase strongly and systematically with spatial scale for all clouds and thresholds, consistent with the near-linear scaling $\tilde{\Lambda} \propto Y$ found in Sect.~\ref{fil_lindens}. On the smallest scales, fractions above $\Lambda_{\rm c}$ are near zero for most clouds, with only Orion\,A (0.03) and Vela\,C (0.05) showing nonzero values on the 14{\arcsec} scale, reflecting the predominantly subcritical nature of small-scale filamentary structures. On the largest scale of 216{\arcsec}, fractions above the nominal threshold range from 14\% in Ophiuchus to 95\% in Vela\,C.

The cloud-to-cloud variation in these fractions is substantial and broadly consistent with the known star formation activity of the clouds. When combined over all spatial scales, the fractions above the nominal threshold $\Lambda_{\rm c}$ range from $\sim$\,7\% in Ophiuchus and $\sim$\,8\% in Taurus to $\sim$\,54\% in Vela\,C, with Perseus (11\%), California (12\%), IC\,5146 (23\%), and Orion\,A (26\%) at intermediate values. At the definitively supercritical threshold $2\Lambda_{\rm c} = 30\,M_\odot$\,pc$^{-1}$, the combined fractions drop substantially: only Vela\,C (35\%), Orion\,A (14\%), and IC\,5146 (12\%) have fractions above 10\%, while all remaining clouds are below 5\%. At the lower threshold $\Lambda_{\rm c}/2 = 7.5\,M_\odot$\,pc$^{-1}$, the combined fractions are considerably higher, ranging from 16\% in Ophiuchus to 71\% in Vela\,C, with all clouds showing fractions above 18\%, suggesting that a significant fraction of filamentary structures in all clouds may be gravitationally unstable within the uncertainties of $\Lambda_{\rm c}$.

The strong scale dependence of these fractions has a direct physical interpretation. Since $\tilde{\Lambda} \propto Y$, larger-scale filaments encompass more mass per unit length and are therefore more likely to exceed $\Lambda_{\rm c}$, regardless of what the true uncertainty in $\Lambda_{\rm c}$ may be. This suggests that gravitational instability and subsequent fragmentation into dense cores develop preferentially in larger-scale, more massive filamentary structures, while smaller-scale filaments remain predominantly subcritical. The results for Vela\,C reported in \cite{Li+2023} provide direct observational support for this picture: supercritical filaments in that cloud contain exclusively prestellar and protostellar cores, with no unbound starless cores, confirming the association between high linear density and active star formation.

Our new estimates of the boundary radius $R$ and the volume density width $h$ have direct implications for theoretical predictions of core and clump separations along filaments. In the sausage instability scenario \citep[e.g.,][]{Jackson+2010}, the expected spacing between fragments is ${\sim\,}11R$ for an incompressible fluid, and ${\sim\,}22H_{\rm th}$ for an infinite isothermal cylinder, where $H_{\rm th}$ is the isothermal scale height. We note that $H_{\rm th}$ characterizes the volume density profile, so it is most naturally compared with our volume density width $h$ rather than the surface density width $H$. For the Ostriker isothermal profile ($\beta = 4$) specifically, $h \approx 3.6\,H_{\rm th}$, so the two are proportional in that limiting case, whereas for shallower profiles the relationship breaks down. In our analysis, both $R$ and $h$ grow systematically with spatial scale (Sects.~\ref{fil_widths} and \ref{discuss_scale_depend}), implying that the predicted fragment separations are themselves scale-dependent rather than fixed. Moreover, for shallow profiles ($\beta \la 2$), the surface density width $H$ substantially overestimates the true physical size $h$ of the filament, so that separation predictions based on $H$ alone would be correspondingly overestimated. We therefore caution that comparisons between observed core separations and analytical instability predictions should account for the scale of observation, the profile shape, and the distinction between surface and volume density widths.

\subsection{Angular resolution and distance effects}
\label{discuss_distance}

Angular resolution strongly affects the results and physical interpretation of filament extractions in star-forming regions \citep[e.g.,] []{Louvet+2021}. Insufficient resolution prevents accurate derivation of their fundamental properties, blending observed structures with nearby ones. A common assumption in previous studies is that observed filaments are well resolved, but this is not always correct -- for shallow profiles, filaments may appear well resolved ($\mathcal{R} \gg 1$) while the true physical filament remains much narrower than the beam \citep[Appendix~A in] []{Menshchikov+Zhang2026subm}.

Comparison of filament properties across our seven molecular clouds is complicated by the wide range of distances, from 140 to 920\,pc. The same angular resolution of $O = 13.5${\arcsec} corresponds to linear scales between 0.009 and 0.06\,pc across this distance range, so that physically similar filaments in more distant clouds suffer substantially greater blending with nearby structures and less accurate background subtraction. Measured properties of blended structures may describe the combined blend rather than individual filaments, potentially biasing comparisons between clouds.

The convergence tests described in Sect.~\ref{convergence_tests} allow us to assess the magnitude of resolution and blending effects independently of physical differences between clouds. Figure~\ref{fig:width_slope_linedens_dist} reveals two unexpected trends with increasing cloud distance: the observed widths increase more steeply with distance than predicted by the isolated filament model, and the observed linear densities also increase with distance, contrary to the expectation that $\Lambda$ should be distance-independent for isolated filaments. The convergence tests (Fig.~\ref{fig:width_slope_linearden_vs_distance}) show that convolving each observed region to progressively lower resolutions -- thereby simulating observations at larger distances while excluding any physical differences between clouds -- reproduces a similar increase in linear densities as observed across clouds at different distances. Since linear density is strictly distance-independent for isolated filaments, as demonstrated in Sect.~\ref{fil_distances}, its increase in the convergence tests can only be attributed to blending with surrounding structures, which adds mass to the measured profiles at lower resolution. This confirms that blending effects alone are sufficient to explain both the anomalous steepening of the width--distance relation and the increase of linear densities with distance, without invoking physical differences between clouds.

However, the observed cloud-to-cloud variation in widths and linear densities is substantially larger and more irregular than predicted by either the isolated filament models or the convergence tests. While models predict a monotonic increase of widths with distance and distance-independent linear densities for isolated filaments, the convergence tests show that blending causes widths to increase more steeply than the models and linear densities to rise by a factor of $\sim$\,4 over a decade in distance. The observed cloud-to-cloud trends are more complex still: the median linear densities drop by a factor of $\sim$\,2 from \object{Taurus} to 
\object{Ophiuchus}, rise by a factor of $\sim$\,4 through \object{Perseus} to \object{Orion\,A}, drop again by a factor of $\sim$\,2--3 for \object{California}, and then a shallower rise to \object{IC\,5146} and a very steep rise to \object{Vela\,C}. This irregular behavior reflects genuine physical differences between the clouds -- in their star formation activity, filament masses, and large-scale environments -- that cannot be attributed to blending effects alone. Disentangling these physical differences from 
resolution and blending effects remains an important challenge for comparative filament studies.

These results have practical implications for filament surveys conducted with a fixed angular resolution across clouds at different distances, such as the \textit{Herschel} Gould Belt Survey \citep{Andre+2010} and the HOBYS program \citep{Motte+2010}. Filament width and slope measurements in more distant clouds are likely affected by resolution bias to a degree that depends on the complexity of the filamentary environment, whereas linear densities -- commonly assumed to be robust -- may also be systematically overestimated due to blending effects. Resolution effects should therefore be explicitly assessed before drawing conclusions about cloud-to-cloud variations in filament properties.

\section{Conclusions}
\label{conclusions}

We have analyzed scale-dependent properties of filamentary structures in seven nearby molecular clouds (\object{Taurus}, \object{Ophiuchus}, \object{Perseus}, \object{Orion A}, \object{California}, \object{IC 5146}, and \object{Vela C}) using the multiscale extraction method \textit{getsf}. Alongside the usual surface density profiles $\Sigma(r)$, we derived volume density profiles $\rho(r)$ for a large sample of filaments, providing new observational constraints on their three-dimensional structure. Unlike previous observational studies that characterized filaments by properties averaged over their entire lengths, our segment-based measurements capture local variations along filament crests, providing a more detailed picture of filament properties and their distributions within and among molecular clouds.

Filament widths $H_k$ and $h_k$ systematically increase with spatial scale, following power laws $\tilde{H} \propto Y^{0.50}$ and $\tilde{h} \propto Y^{0.37}$ across all clouds, with width distributions spanning $\sim$\,0.01--0.1\,pc on the smallest 14{\arcsec} scale and $\sim$\,0.01--1\,pc on the largest 216{\arcsec} scale. These results challenge the notion of a universal filament width of $\sim$\,0.1\,pc. The median volume density slopes $\tilde{\beta} \approx 2.1$--$2.4$ are systematically lower than the value $\beta = 4$ expected for an isothermal cylinder in hydrostatic equilibrium, indicating that observed filaments are generally not well described by this idealized model. For profiles with $\beta \la 2$, the volume density width $h$ is systematically smaller than the surface density width $H$, with the difference strongly increasing toward shallower slopes and reaching one to two orders of magnitude for $\beta \la 1$, demonstrating that surface density widths overestimate the true physical width of filaments with shallow profiles. The volume density contrasts $C_{\rho}$ are substantially higher than the surface density contrasts $C_{\Sigma}$, with median values of $\tilde{C}_{\rho} \approx 17$--$52$ compared to $\tilde{C}_{\Sigma} \approx 1.1$--$2.7$ across all scales and clouds, confirming that filaments are substantially more prominent in three dimensions than their projected appearance suggests.

The median linear densities derived from all seven clouds linearly increase with the spatial scale, $\tilde{\Lambda} \propto Y$, with values spanning [0.07, 3.4]\,$M_\odot$\,pc$^{-1}$ on the 14{\arcsec} scale and [3.2, 63]\,$M_\odot$\,pc$^{-1}$ on the 216{\arcsec} scale. The fraction of supercritical filaments ($\Lambda > 15\,M_\odot$\,pc$^{-1}$) strongly increases with the spatial scale for all clouds, varying widely among clouds from $\sim$\,7\% in Taurus and Ophiuchus to $\sim$\,54\% in Vela C when combined over all spatial scales, which is broadly consistent with the known star formation activity of the clouds. The transition from subcritical to supercritical behavior with an increasing scale suggests that gravitational instability develops, preferentially, in larger-scale, more massive filamentary structures. The statistical properties of the combined filament linear density function across all seven clouds, including its connection to the stellar initial mass function, will be presented in a subsequent paper.

Measured filament widths and slopes systematically depend on angular resolution and distance. Convergence tests confirm that resolution degradation increases measured widths and slopes, and this also raises measured linear densities due to blending with surrounding structures, though the increase in linear densities is less steep with resolution degradation than the observed cloud-to-cloud trend suggests. The irregular cloud-to-cloud variation in widths and linear densities reflects genuine physical differences between the clouds that cannot be attributed to blending effects alone.

\begin{acknowledgements}
Images of the molecular clouds were downloaded from the \textit{Herschel} Science Archive\footnote{\url{http://archives.esac.esa.int/hsa/whsa/}} and obtained in the \textit{Herschel} Gould Belt Survey (HGBS\footnote{\url{http://gouldbelt-herschel.cea.fr}}) (PI: Ph.\,Andr{\'e}), HOBYS\footnote{\url{http://hobys-herschel.cea.fr}} (PI: F.\,Motte, A.\,Zavagno, S.\,Bontemps), and A-CMC (PI: P.\,Harvey). HGBS and HOBYS are the \textit{Herschel} Key Projects jointly carried out by SPIRE Specialist Astronomy Group 3 (SAG3), scientists of several institutes in the PACS Consortium (e.g., CEA Saclay, INAF-IAPS Rome, LAM/OAMP Marseille), and scientists of the \textit{Herschel} Science Center (HSC). 

This work was supported by the Key Project of International Cooperation of the Ministry of Science and Science of China through grant 2010DFA02710, and by the National Natural Science Foundation of China through grants 11503035, 11573036, 11373009, 11433008, 11403040, and 11403041. G.Z. acknowledges support from the Postdoctoral Science Foundation of China (No. 2021T140672). 
The authors thank the anonymous referee for constructive comments that prompted improvements to this paper.
\end{acknowledgements}

\bibliographystyle{aa}
\bibliography{filament_scale_dependency}

\begin{appendix}

\onecolumn
\section{Accuracy of skeletons and filament measurements}
\label{app:accskeletons}

\begin{figure}[ht!]
\centering
\includegraphics[width=0.6\hsize]{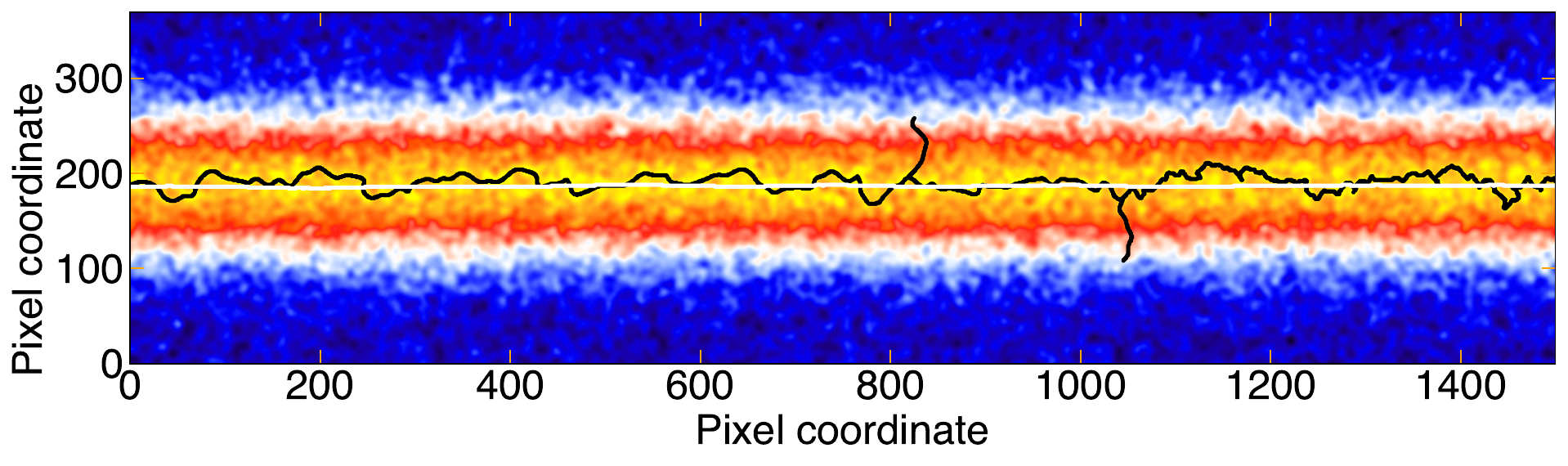}
\caption{
Comparison of scale-dependent skeletons traced by \textit{getsf} (white line) and \textit{disperse} (black line) for a simulated straight filament. The image has an angular resolution of 8{\arcsec} and pixel size of 2{\arcsec}. The simulated filament has a true half-maximum width of 128{\arcsec} and random small-scale unresolved fluctuations corresponding to a crest-to-noise ratio of 20. For fainter filaments (with lower crest-to-noise ratios), the \textit{disperse} skeletons become even more wiggly. This irregular behavior of the \textit{disperse} skeletons is also found for similar filaments with narrower widths of 64, 32, and 16{\arcsec}, though to a progressively lesser extent; this irregular behavior practically disappears only for unresolved filaments. The \textit{disperse} skeleton shown here was derived with the following free parameters: persistence threshold 40, robustness threshold 60, and smoothing length 30 pixels. Without the artificial smoothing, the skeleton is even more irregular. The \textit{getsf} skeletons are unaffected by the fluctuations and were detected as practically straight lines for all filament widths on all spatial scales.
}
\label{fig:disperse_overlay}
\end{figure}

Skeletons trace locations of filament crests and define the orthogonal directions (normals) along which the filament profiles are obtained, determining fundamental physical properties of the filaments (widths, slopes, linear densities, masses, and other properties). Therefore, the accuracy of the skeletons is important not only for visualizing filament locations, but also for accurate measurements. However, the standard approach of tracing skeletons directly in observed images becomes highly inaccurate for resolved filaments because of the presence of fluctuations on all spatial scales. Especially problematic is the presence of sources and small-scale fluctuations produced by either background or instrumental noise. The fluctuations on scales on the order of the angular resolution are strong enough to alter the shapes of the filament crests. Likewise, sources of various sizes embedded in filaments have peaks that disturb the filament crest shapes. This is why the standard approach can lead to inaccurate, wiggly skeletons (Fig.~\ref{fig:disperse_overlay}) and flawed measurements, especially for more resolved and lower-contrast filaments.

The multiscale approach followed by \textit{getsf} solves this problem. Instead of extracting sources and filaments in observed images, the method analyzes a large set of decomposed single-scale images \citep[Fig.~7 in] []{Menshchikov2021method}. Each decomposed image contains structures from only a limited range of spatial scales, so that contributions of small-scale fluctuations are effectively filtered out in larger-scale decomposed images. Moreover, \textit{getsf} extracts sources before extracting filaments, and thus completely removes round small-scale peaks from the original images \citep[Fig.~8 in] []{Menshchikov2021method}. Furthermore, the scale-dependent detection of skeletons is done within a narrow range of spatial scales, where contributions from much smaller and larger scales are removed. This ensures that the skeletons trace the filament crests much better, even for wider and fainter filaments \citep[Fig.~13 in] []{Menshchikov2021method}. As demonstrated in Fig.~\ref{fig:disperse_overlay}, the \textit{getsf} skeletons accurately trace the straight filament crest, whereas the skeletons produced by \textit{disperse} are quite wiggly, leading to normals that strongly deviate from the correct orthogonal directions of measurements for the simulated filament.

\clearpage

\section{Scale-dependent skeletons in molecular clouds}
\label{app:skeletons}

This appendix presents images of the scale-dependent skeletons detected in all seven molecular clouds.

\begin{figure*}[ht!]
\centering
\includegraphics[width=1.0\hsize]{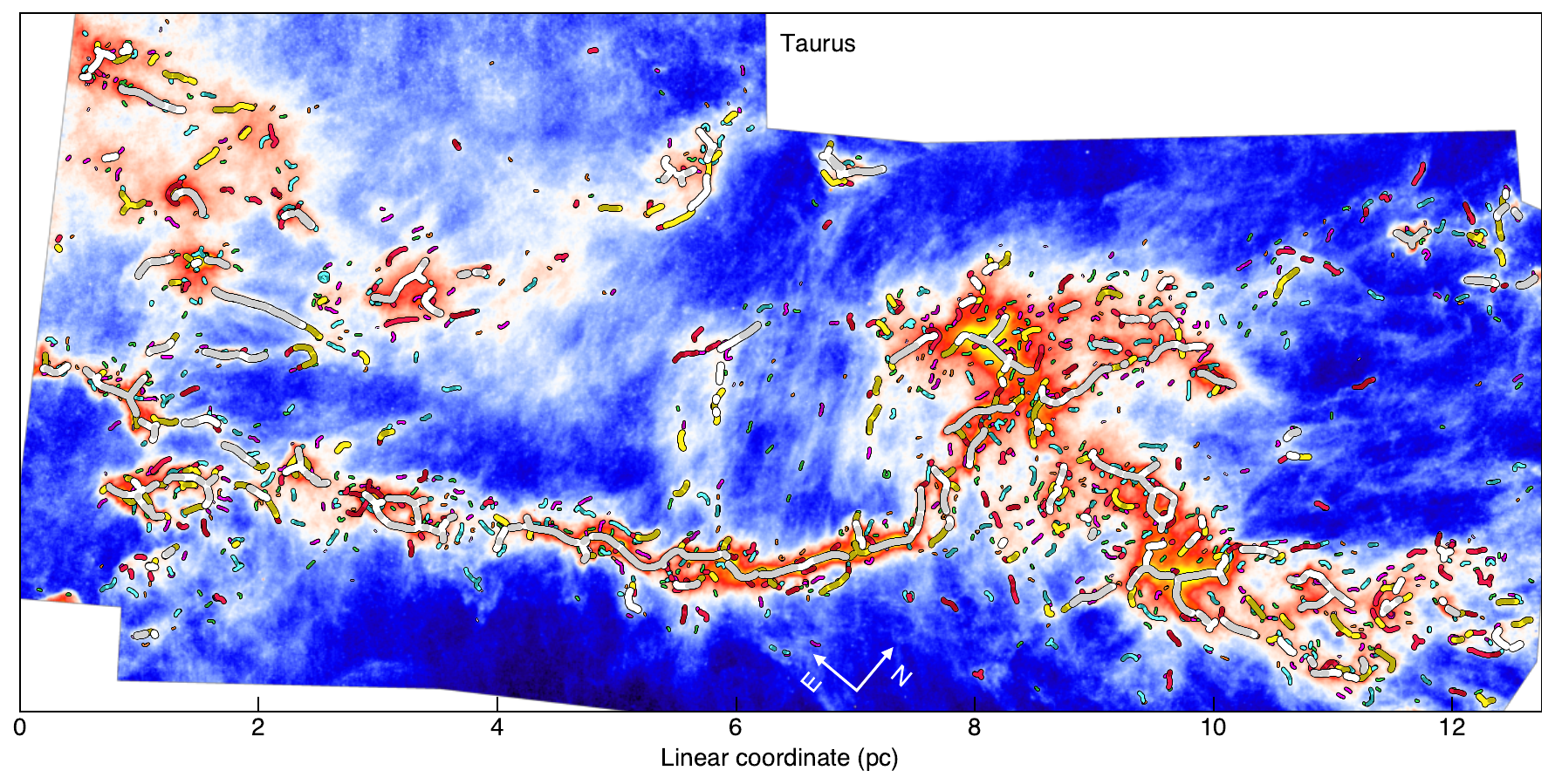}\\
\includegraphics[width=1.0\hsize]{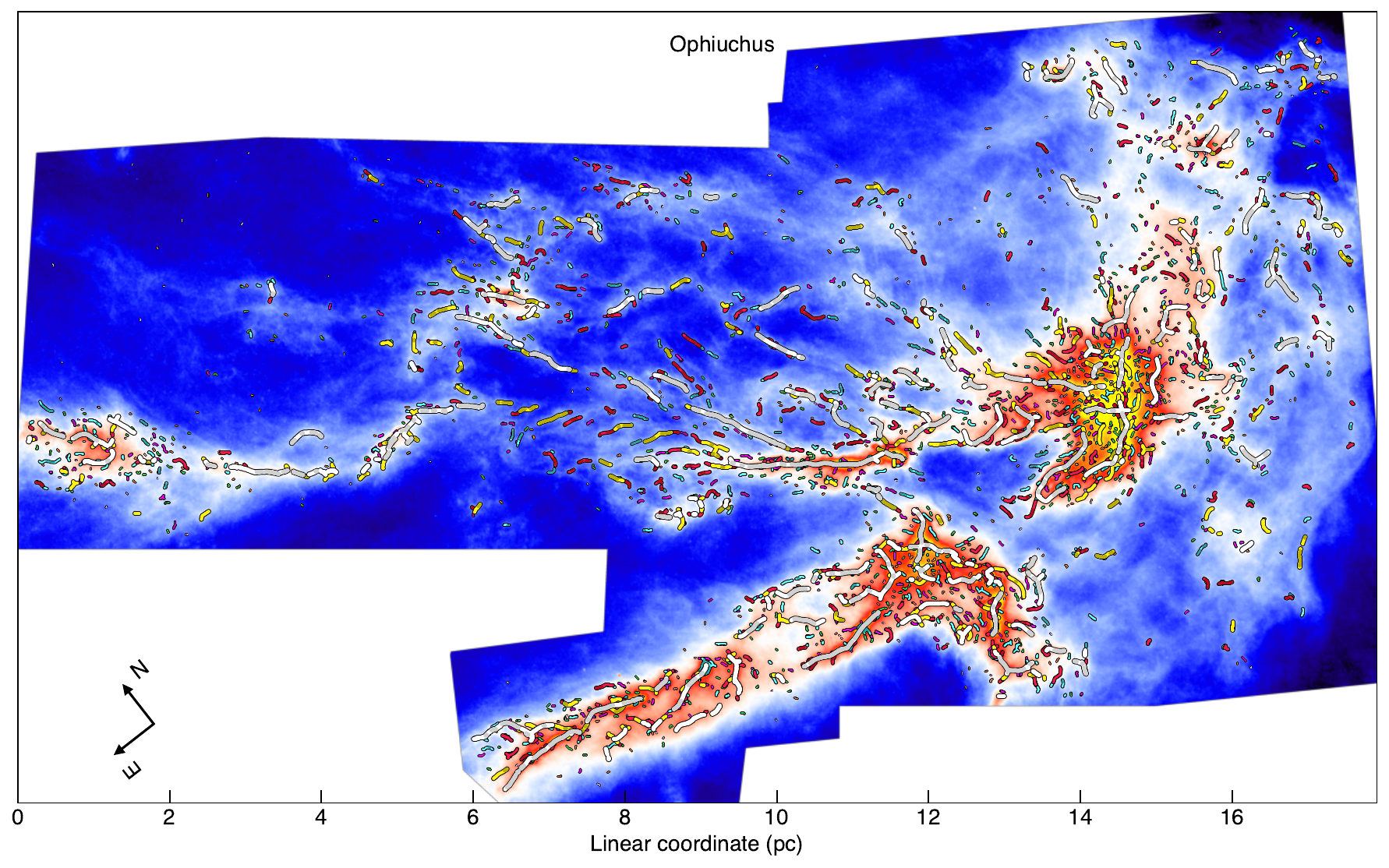}\\
\includegraphics[width=1.0\hsize]{filament_scale_dependency_figs/Taurus.r13p5.surfacedensity.skeleton.distinct.colorbar.pdf}
\caption{
Surface density images of molecular clouds overlaid with scale-dependent skeletons traced by \textit{getsf} on scales 14--216{\arcsec}. This figure shows \object{Taurus} (top) and \object{Ophiuchus} (bottom). Skeleton widths are proportional to the detection scale, as indicated by the color bar. Darker colors indicate segments with profiles meeting the quality criteria (Sect.~\ref{fil_select}). The color bar applies to all panels in this figure and Figs.~\ref{fig:filament.skeleton.clouds.b}--\ref{fig:filament.skeleton.clouds.c}. See also Fig.~\ref{fig:filament.skeleton.clouds.b} (\object{Perseus}, \object{Orion A}, \object{California}) and Fig.~\ref{fig:filament.skeleton.clouds.c} (\object{IC 5146}, \object{Vela C}).
}
\label{fig:filament.skeleton.clouds.a}
\end{figure*}

\begin{figure*}[ht!]
\centering
\includegraphics[width=1.0\hsize]{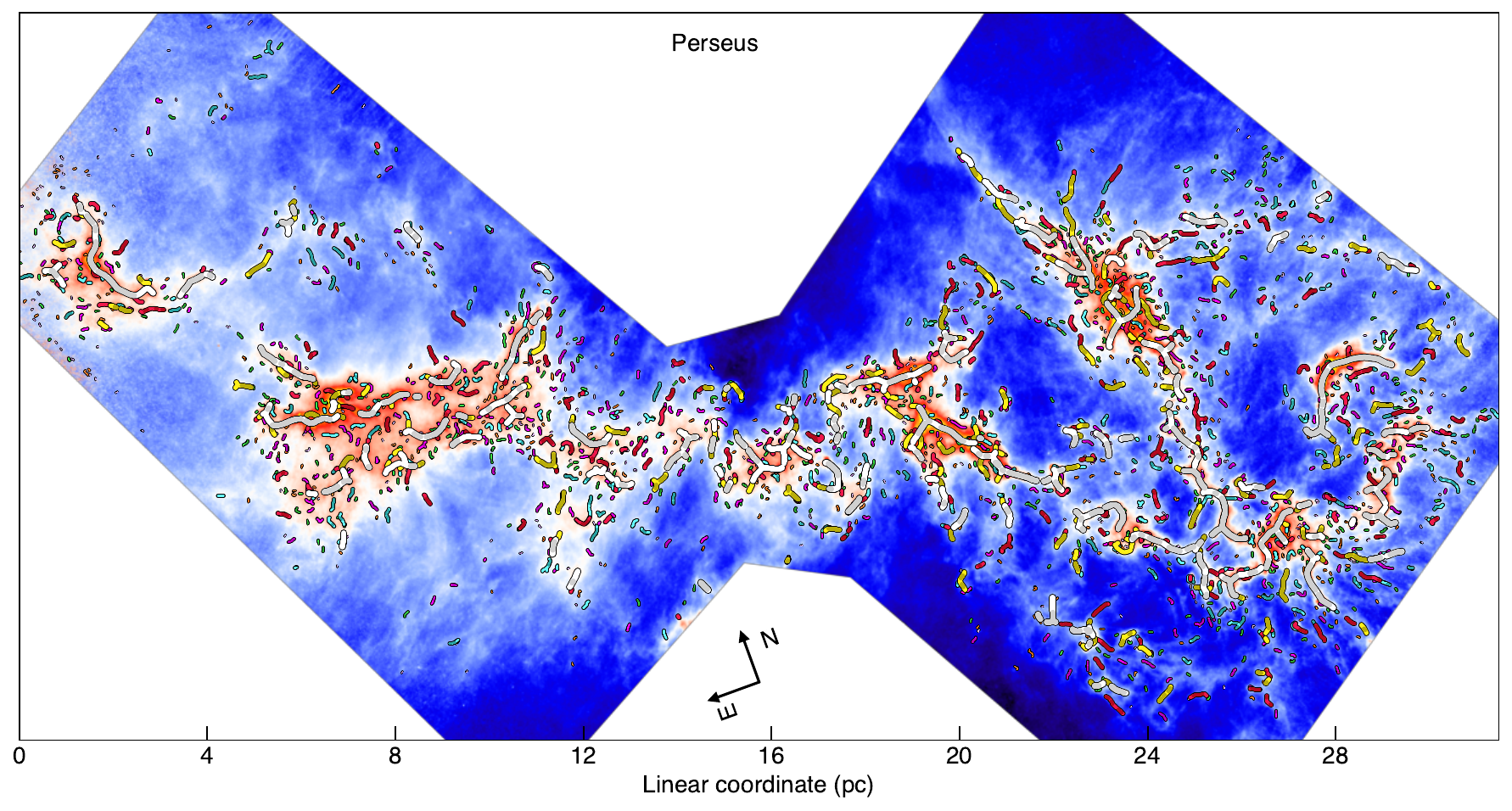}\\
\includegraphics[width=1.0\hsize]{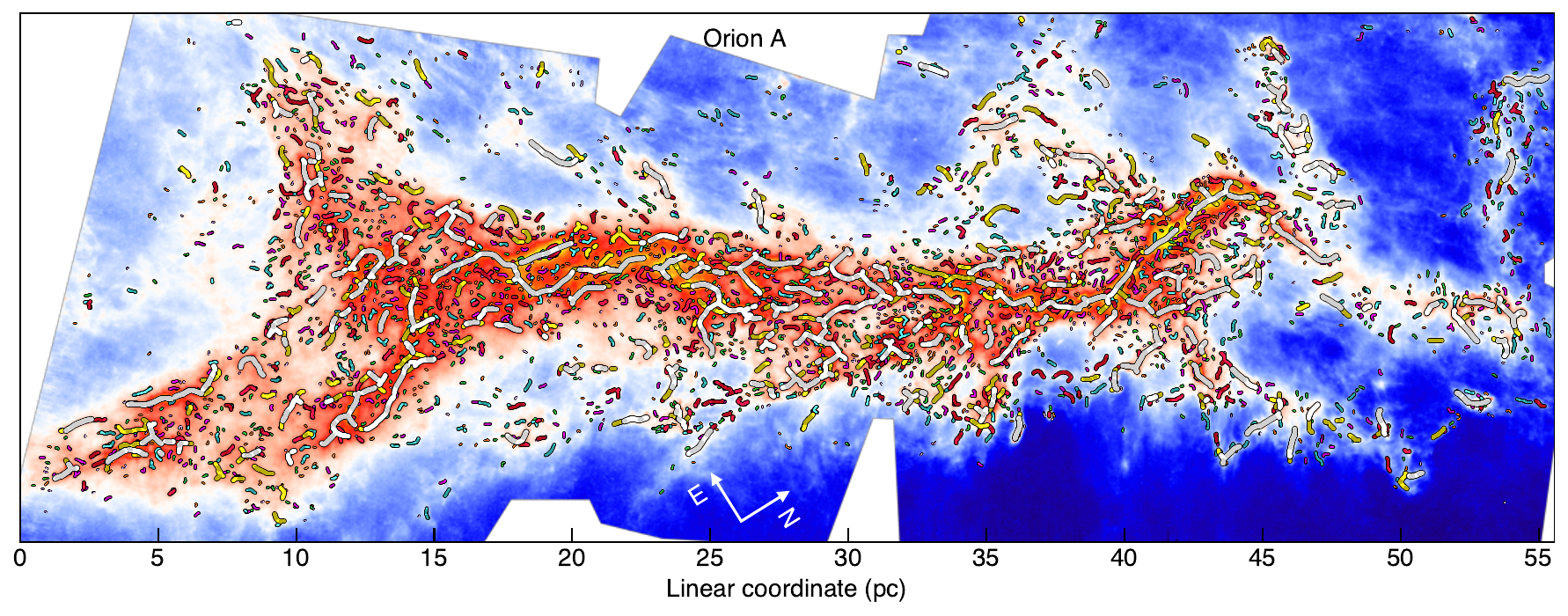}\\
\includegraphics[width=1.0\hsize]{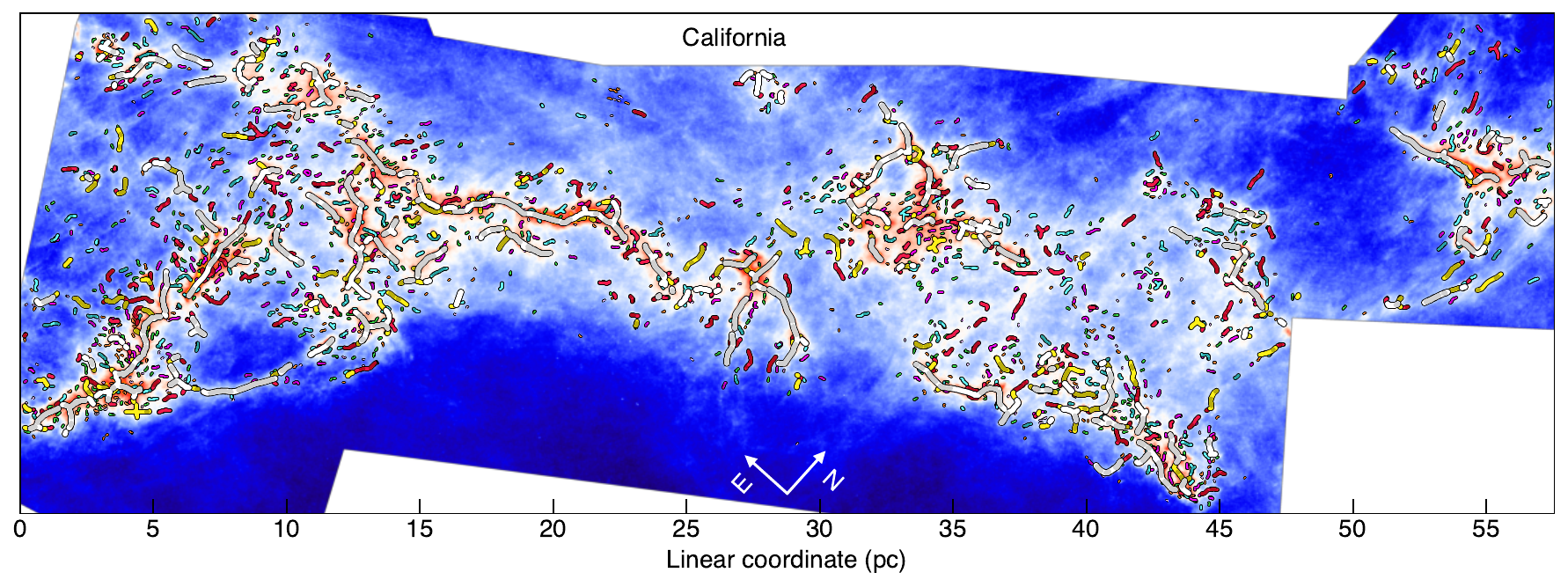}
\caption{
Continued from Fig.~\ref{fig:filament.skeleton.clouds.a}. From top to bottom: \object{Perseus}, \object{Orion A}, and \object{California}. See Fig.~\ref{fig:filament.skeleton.clouds.c} for \object{IC 5146} and \object{Vela C}.
}
\label{fig:filament.skeleton.clouds.b}
\end{figure*}

\begin{figure*}[ht!]
\centering
\includegraphics[width=1.0\hsize]{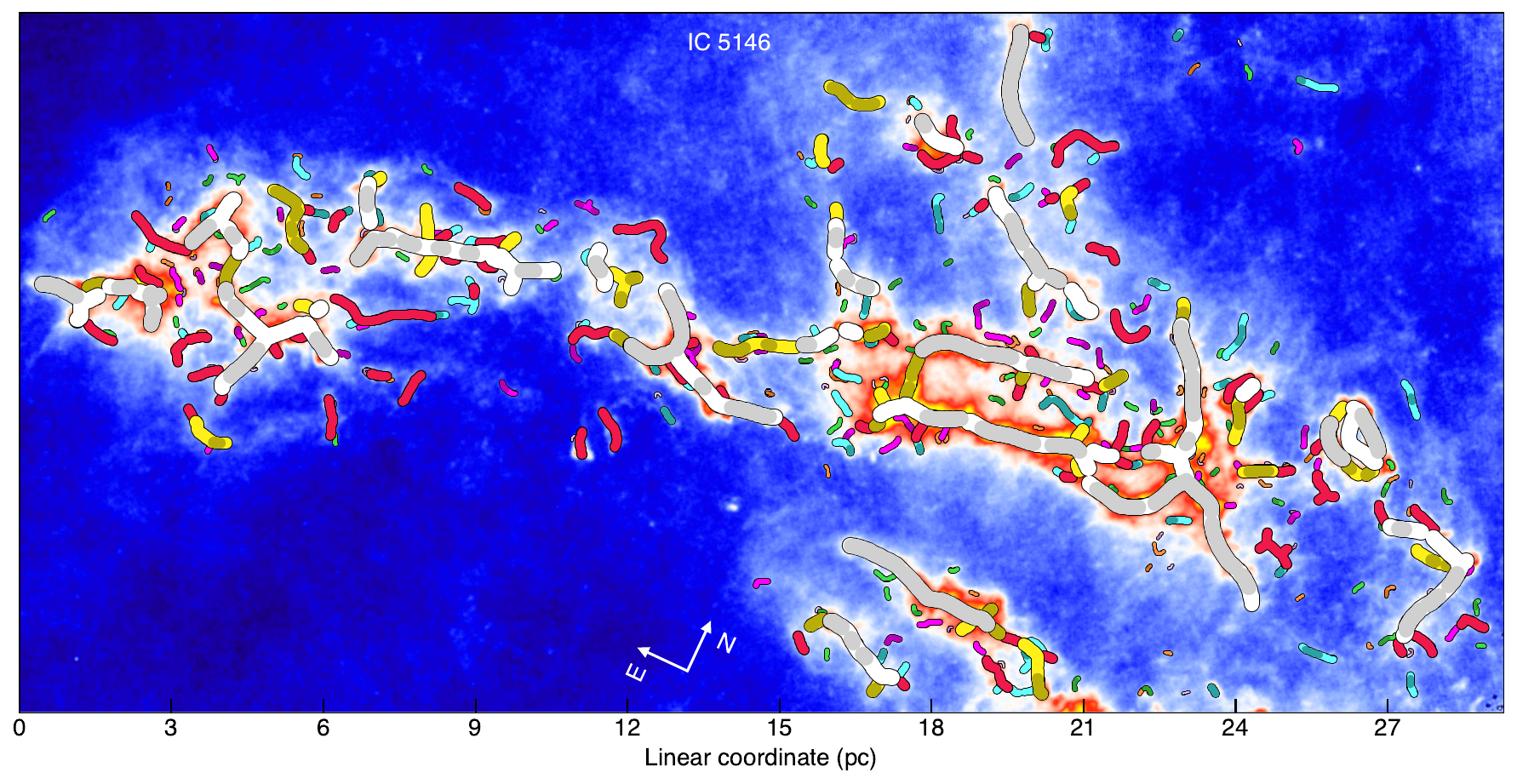}\\
\includegraphics[width=1.0\hsize]{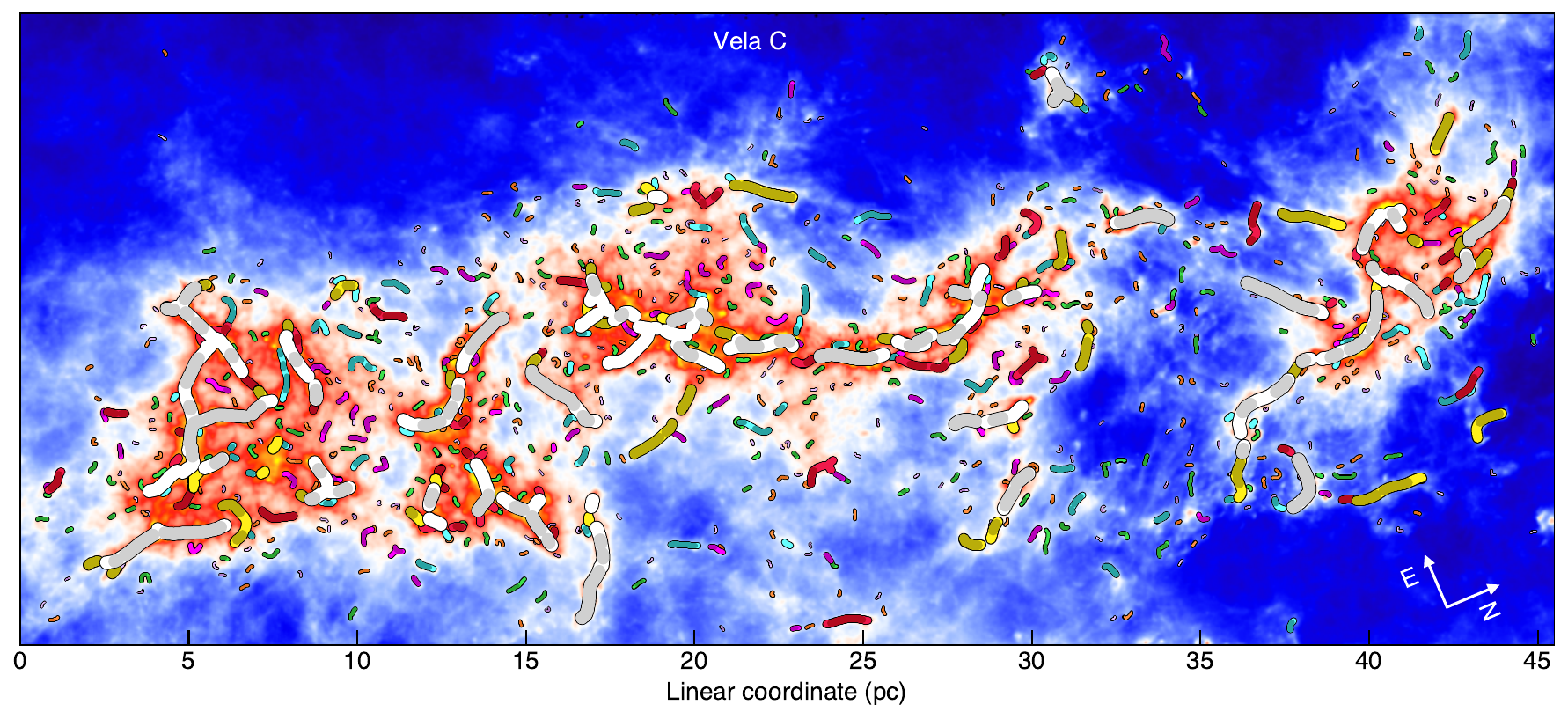}
\caption{
Continued from Figs.~\ref{fig:filament.skeleton.clouds.a} and \ref{fig:filament.skeleton.clouds.b}. From top to bottom: \object{IC 5146} and \object{Vela C}.
}
\label{fig:filament.skeleton.clouds.c}
\end{figure*}

\FloatBarrier 

\section{Surface density profiles of filaments}
\label{app:surfdensprofs}

\begin{figure}[ht!]
\centering
\includegraphics[width=0.5\hsize]{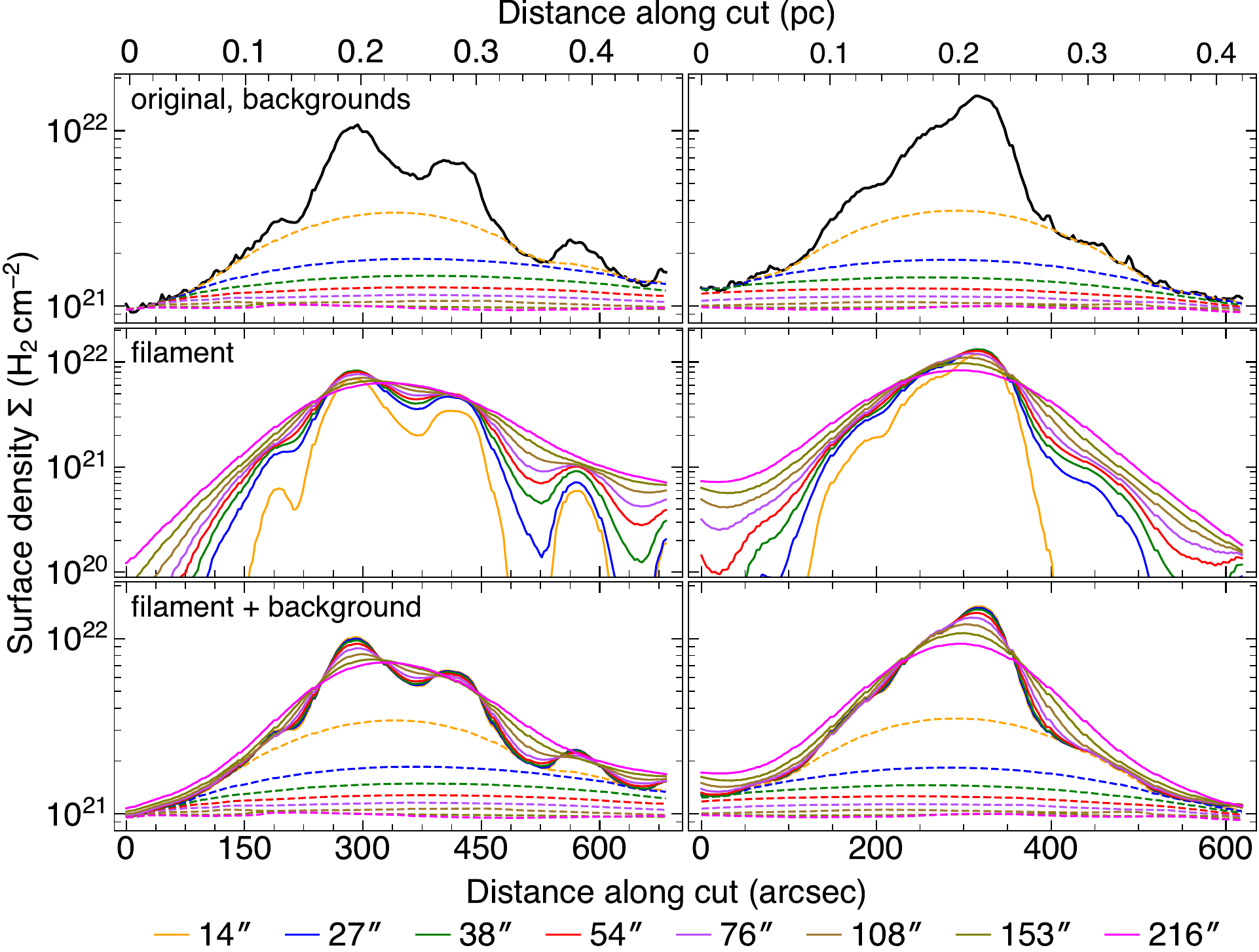}
\caption{
Surface density profiles along two parallel cuts across the \object{Taurus} filament (see Fig.~\ref{fig:filament.skeleton.taurus}). \textit{Top panels:} Observed filament profiles (solid curves) and scale-dependent backgrounds (dashed curves) derived by \textit{getsf} on each spatial scale. \textit{Middle panels:} Scale-dependent filament profiles obtained by subtracting the backgrounds from the observed profiles, with contributions from smaller scales removed. \textit{Bottom panels:} Smooth profiles (solid lines) obtained by summing the scale-dependent backgrounds and filaments on each spatial scale, with background profiles (dashed, from top panels) shown for reference.
}
\label{fig:taurus_cut_profiles}
\end{figure}

Figure~\ref{fig:taurus_cut_profiles} illustrates the complex structure and substructure of observed filaments. Several peaks in the filament profiles clearly visible in the middle panels indicate that there are distinct blended substructures with substantially different widths within the large Taurus filament. These structures are detected by \textit{getsf} using scale-dependent skeletons tracing the smooth crests of the filaments and sub-filaments.

\begin{figure}[ht!]
\centering
\includegraphics[width=0.5\hsize]{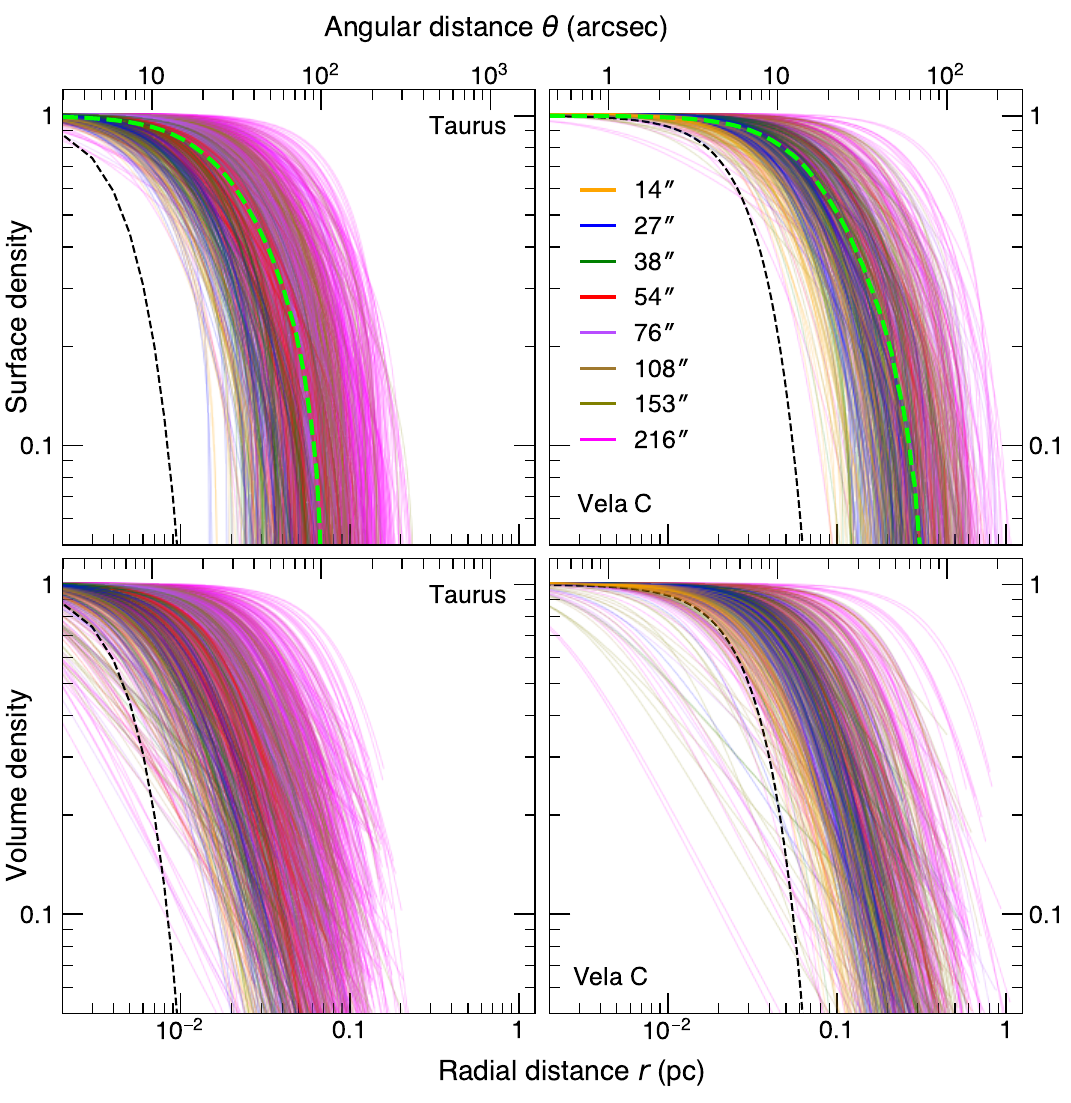}
\caption{
Normalized surface and volume density profiles $\Sigma(r)$ and $\rho(r)$ for the nearest and furthest molecular clouds chosen to illustrate the range of angular resolution and distances in our sample. Colored curves represent profiles from Eqs.~(\ref{volume_density}) and (\ref{surfdens_fittingfun}) for filament segments detected on spatial scales 14--216{\arcsec}. Green dashed lines show the median surface density profiles over all spatial scales. For reference, short-dashed curves show the 13.5{\arcsec} beam profile. Top axes show the angular scales corresponding to the physical scales at the distance of each cloud.
}
\label{fig:filament_radial_profiles_fitted_surface_volume}
\end{figure}

Figure~\ref{fig:filament_radial_profiles_fitted_surface_volume} shows the scale-dependent profiles $\Sigma_{k}(r)$ of each filament segment and the median profiles $\tilde{\Sigma}_{k}(r)$ over all segments on scale $Y_k$. For clarity, the individual $\Sigma_{k}(r)$ profiles were plotted only up to their truncation points $r_{\rm T}$.

\section{Analytical relationships between volume and surface density}
\label{app:fittingfun}

\cite{Menshchikov+Zhang2026subm} developed a fitting method in which the geometry-induced radial dependence of surface density profiles $\Sigma(r)$ of finite cylindrical structures is explicitly taken into account and the profiles are related to and consistent with volume density profiles $\rho(r)$. We briefly summarize here the analytical formulae relevant for fitting observed profiles using Eq.~(\ref{surfdens_fittingfun}). These are numerical approximations derived and validated in \cite{Menshchikov+Zhang2026subm}, to which we refer for full details.

We assume that the volume density distribution within radius $R$ of the filament circular boundary can be approximated by
\begin{equation}
\rho(r) = \rho_{\rm C}\left(1+\left(2^{2/\beta\!}-1\right)\left(\frac{2r}{h}\right)^2\right)^{-\beta/2}\,,
\label{volume_density}
\end{equation} 
where $\rho_{\rm C}$ is the axial volume density, $h$ is the full half-maximum width of the profile, and the factor $(2^{2/\beta}-1)$ ensures that $h$ is the half-maximum width of the profile independently of $\beta$ \citep{Menshchikov2021method}. The slope $\beta$ is related to the intrinsic slope $\gamma$ and the extent ratio $\xi$ by
\begin{equation}
\beta = \gamma + 1.529 \left(1 + \xi^{-0.199}\exp\left(-2.725 \left(0.5\gamma\left(\xi^{0.03\gamma^{-0.7\!}\!} + \xi^{0.26\gamma^{-0.1}\xi^{0.03}}\right) -0.319\right)\right)\right)^{-1} - 0.541,
\label{betaformula}
\end{equation}
where $\xi \equiv R/h$ is the extent ratio of the power-law envelope. The empirical exponent $\epsilon$ in Eq.~(\ref{surfdens_fittingfun}) ensures consistency of $\Sigma(r)$ toward the boundary at $r\rightarrow R$ with the $\rho(r)$ profile in Eq.~(\ref{volume_density}) and can be written as
\begin{equation}
\epsilon = 40.115 \left(1 + \exp\left(-0.3782 \left(\beta + 2.542\right)\right) \xi^{0.012}\right)^{-1} + 8.21 \exp\left(-0.7497 \beta \xi^{0.0475}\right) \xi^{-0.015} - 34.104.
\label{epsilonformula}
\end{equation}
The intrinsic half-maximum width $w$ of the surface density profile can be found from $\beta$, $\xi$, and the measured width $H$ as
\begin{equation}
w = H \left(235.7\exp\left(-20\beta \xi^{-0.2}\right) + 0.00005\xi^{0.5}\beta^{-6\xi^{0.21\!}\!}+2.878\exp\left(-1.069\beta \xi^{0.22\beta}\right) + 113.0\exp\left(-10.88\beta \xi^{-0.2}\right)+ 1.022\xi^{-0.0077} \right).
\label{wHformula}
\end{equation}
The half-maximum width $h$ of the corresponding volume density profile can also be derived from $\beta$, $\xi$, and the width $H$ as
\begin{equation}
h = H \left(S+\left(E-S\right)\left(1+\left(\beta/G\right)^{F}\right)^{-Z}\right),
\label{hformula}
\end{equation}
where the coefficients $E$, $F$, $G$, $S$, and $Z$ are functions of $\xi$:
\begin{align}
\begin{split}
E &= 0.77149\left(1-\exp\left(-\left(\left(\xi-0.52709\right)/0.7156\right)^{-0.9095}\right)\right)+0.0026857, \\
F &= -0.31586\xi^{-1.9388\!}+0.57344\xi^{-0.96778\!}+6.5472\xi^{0.14471} - 5.1551, \\
G &= 1.0811\xi^{-0.62466\!}-1.4375\xi^{-1.0203\!}+4.0626\xi^{0.00955} - 3.1165, \\
S &= 0.034613 \exp\left(-0.014394\xi\right) +0.036328\exp\left(-0.27696\xi\right)+2.1537\exp\left(-5.104\xi\right) + 0.94275, \\
Z &= 0.26355 \exp\left(-0.059635\xi\right)+8.8497\exp\left(-4.824\xi\right) -645.57\exp\left(-12.634\xi\right) + 0.24014.
\label{coeffshH}
\end{split}
\end{align}

The fitting function of Eq.~(\ref{surfdens_fittingfun}) has only three independent parameters to be optimized: the intrinsic slope $\gamma$, the boundary radius $R$, and the crest surface density $\Sigma_{\rm C}$ (as noted in Sect.~\ref{fil_measure}). The true extent ratio $\xi$ of the power-law envelope is computed using the bisection method, since $\xi$ appears implicitly on both sides of the equation, after substituting $\beta(\gamma,\xi)$ from Eq.~(\ref{betaformula}) into Eq.~(\ref{hformula}). With $\xi$ determined this way, the parameters $\beta$, $\epsilon$, $w$, and $h$ are computed from Eqs.~(\ref{betaformula})--(\ref{coeffshH}), ensuring full consistency between the volume and surface density profiles.

\section{Scatter plots of filament properties}
\label{app:scatterplots}

\begin{figure*}[ht!]
\centering
\includegraphics[width=\textwidth]{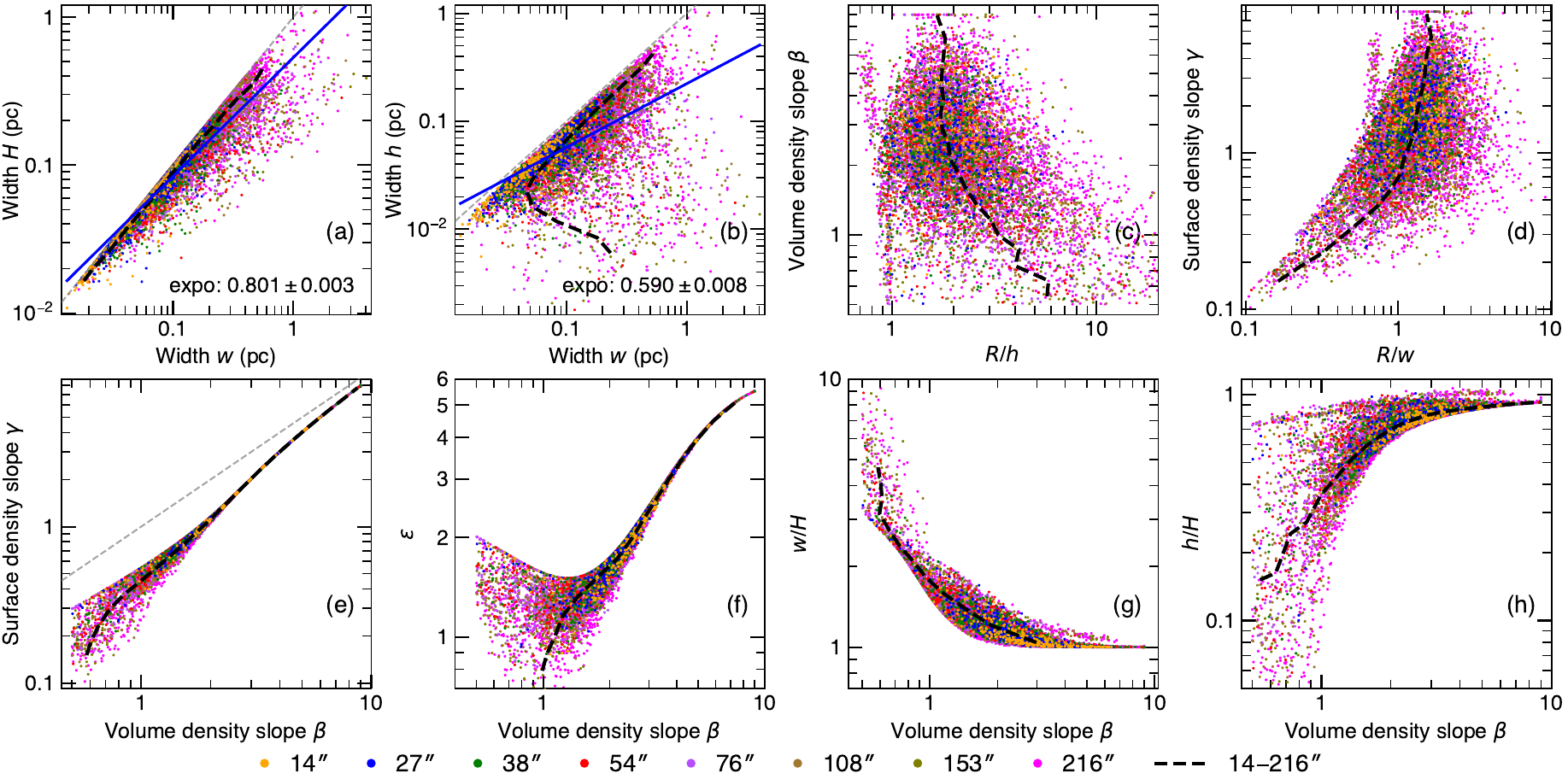}
\caption{
Filament width and density slope relations across molecular clouds (see Appendix~\ref{app:fittingfun} for definitions of all quantities). Panels show:
(\textit{a}) measured width $H$ versus intrinsic width $w$ (from surface density profiles),
(\textit{b}) volume density width $h$ versus intrinsic width $w$,
(\textit{c}) volume density slope $\beta$ versus profile extent $\xi \equiv R/h$,
(\textit{d}) surface density slope $\gamma$ versus ratio $R/w$,
(\textit{e}) surface density slope $\gamma$ versus volume density slope $\beta$,
(\textit{f}) geometry exponent $\epsilon$ versus volume density slope $\beta$,
(\textit{g}) width ratio $w/H$ versus volume density slope $\beta$, and
(\textit{h}) width ratio $h/H$ versus volume density slope $\beta$.
Colored points represent measurements at different spatial scales (14--216{\arcsec}), with black dashed lines showing binned median trends across all scales. Only the $\gamma$ values satisfying our quality criteria (Sect.~\ref{fil_select}) are used in panels (\textit{c})--(\textit{h}). Gray dashed lines in panels (\textit{a}), (\textit{b}), and (\textit{e}) indicate equality relations for reference. Blue solid lines in panels (\textit{a}) and (\textit{b}) represent power-law fits to individual data points with exponents indicated. Distributions of points in panels (\textit{e})--(\textit{h}) reflect the corresponding relations of the fitting model \citep[Fig.~2 in] []{Menshchikov+Zhang2026subm}.
}
\label{fig:combined_width_relations}
\end{figure*}

\begin{figure*}[ht!]
\centering
\includegraphics[width=0.95\textwidth]{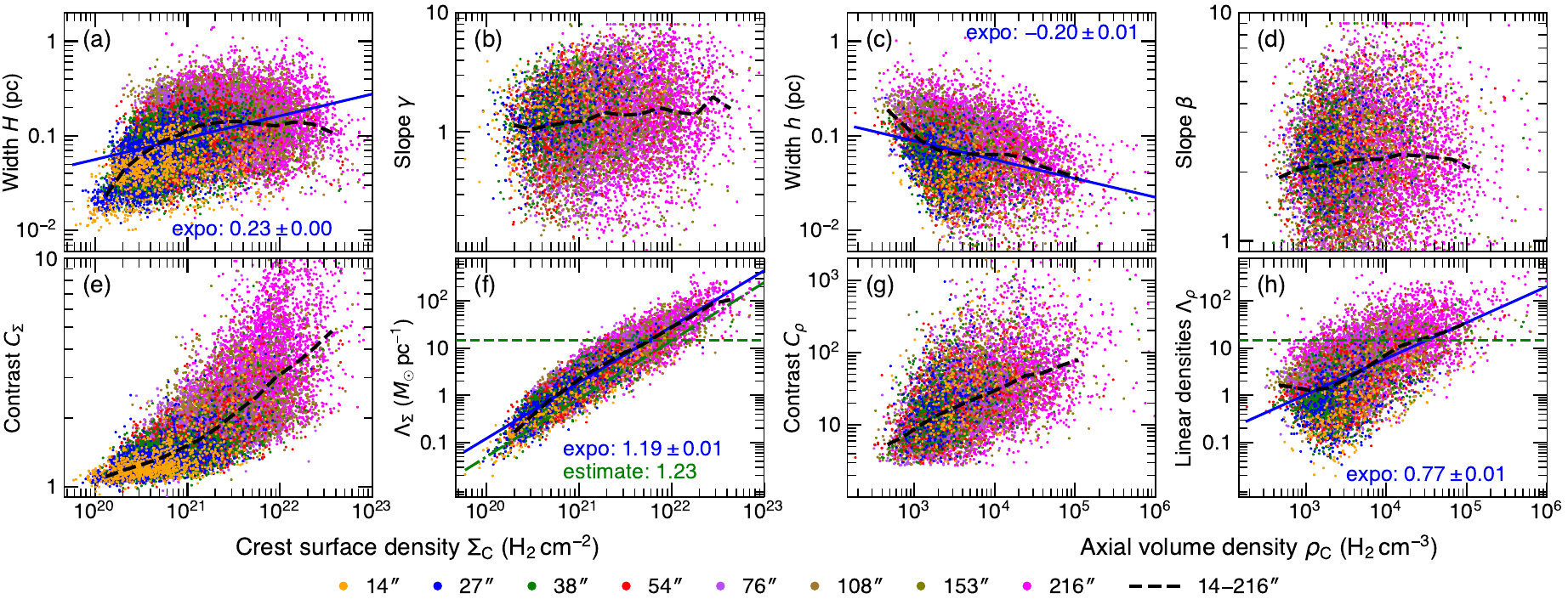}
\caption{
Correlations between filament properties and crest surface density $\Sigma_{\mathrm{C}}$ (left column) and axial volume density $\rho_{\mathrm{C}}$ (right column) across seven molecular clouds. Panels show: (\textit{a}) surface density width $H$, (\textit{b}) intrinsic slope $\gamma$, (\textit{c}) volume density width $h$, (\textit{d}) slope $\beta$, (\textit{e}) surface density contrast $C_\Sigma$, (\textit{f}) surface linear density $\Lambda_\Sigma$, (\textit{g}) volume density contrast $C_\rho$, and (\textit{h}) volume linear density $\Lambda_\rho$. Data points are color-coded by spatial scale (14--216{\arcsec}) as shown at bottom. Blue lines represent power-law fits to the data in panels (\textit{a}), (\textit{c}), (\textit{f}), and (\textit{h}), with the fitted exponent indicated in the panels. Black dashed lines show binned median trends (bins containing $>30$ points). In panel (\textit{f}), the green dot-dashed line illustrates the expected relation $\Lambda_\Sigma \propto \Sigma_{\mathrm{C}} H(\Sigma_{\mathrm{C}})$ derived from the power-law fit in panel (\textit{a}). The green horizontal dashed lines in panels (\textit{f}) and (\textit{h}) mark a reference linear density of $15\,M_\odot\,\mathrm{pc}^{-1}$.}
\label{fig:combined_filament_properties}
\end{figure*}

\begin{figure*}[ht!]
\centering
\includegraphics[width=0.95\textwidth]{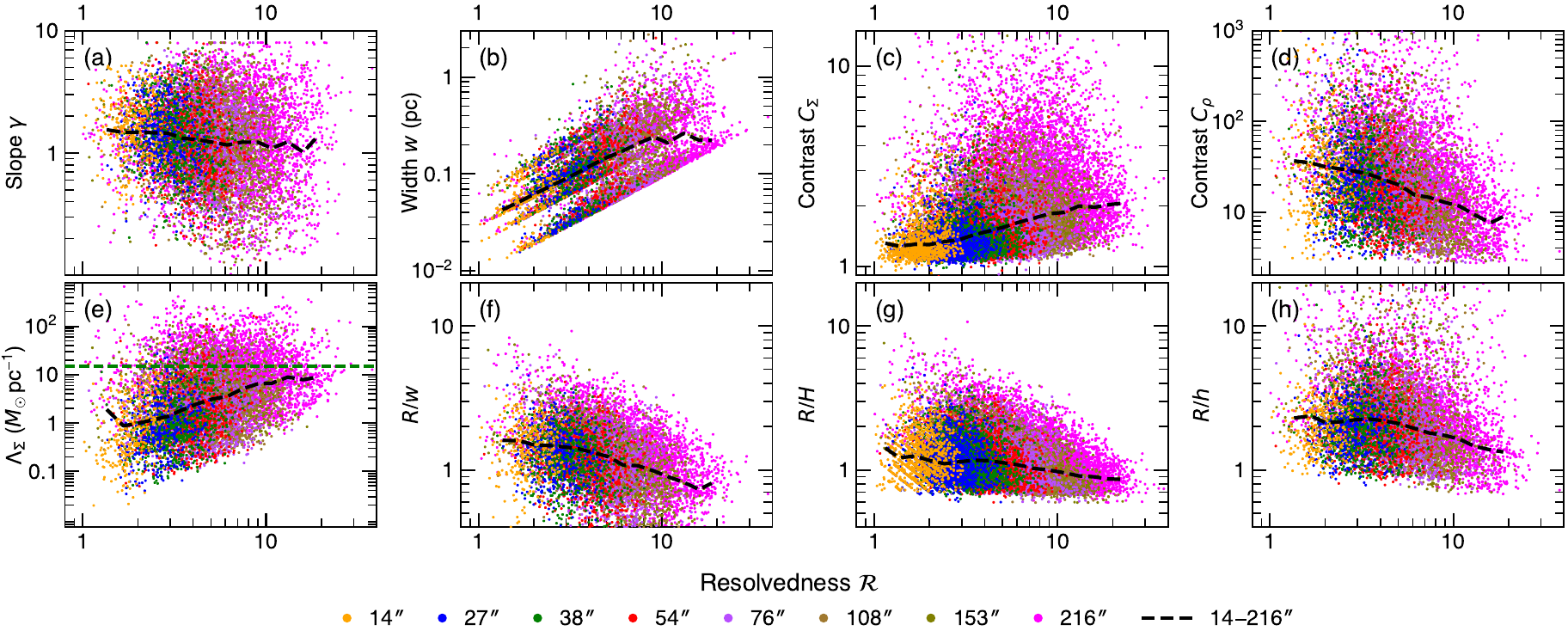}
\caption{
Filament properties as functions of resolvedness $\mathcal{R} \equiv H/O$ for all molecular clouds combined. Panels show: (\textit{a}) surface density slope $\gamma$ (median trend calculated from data with $\gamma < 8$), (\textit{b}) surface density width $w$, (\textit{c}) surface density contrast $C_\Sigma$, (\textit{d}) volume density contrast $C_\rho$, (\textit{e}) linear density $\Lambda_\Sigma$, (\textit{f}) ratio $R/w$, (\textit{g}) ratio $R/H$, and (\textit{h}) profile extent $\xi \equiv R/h$. Black dashed lines show binned median trends across all spatial scales (14--216{\arcsec}).
}
\label{fig:all_properties_vs_resolvedness}
\end{figure*}

This appendix presents correlations among filament properties across all spatial scales and molecular clouds. These plots complement the scale-dependent distributions presented in the main text and illustrate the interrelations among the measured filament parameters. Figure~\ref{fig:combined_width_relations} visualizes key relations between width and density slope parameters, fully consistent with the analytical fitting model described in Appendix~\ref{app:fittingfun} \citep[see also Fig.~2 in] []{Menshchikov+Zhang2026subm}. Figure~\ref{fig:combined_filament_properties} displays filament properties as functions of crest surface density $\Sigma_{\mathrm{C}}$ and axial volume density $\rho_{\mathrm{C}}$, with power-law fits overlaid. Figure~\ref{fig:all_properties_vs_resolvedness} shows how the properties depend on resolvedness $\mathcal{R}$, revealing systematic trends with angular resolution. Collectively, these scatter plots provide a comprehensive view of the interrelations among filament parameters and confirm the internal consistency of the measurements.

\end{appendix}
\end{document}